\documentclass[12pt,a4paper]{article}

\usepackage{setspace}

\usepackage{natbib}
\usepackage{graphics}
\usepackage{enumerate}
\usepackage{graphicx}
\usepackage{amsfonts}
\usepackage{amsmath}
\usepackage{amssymb}
\usepackage{rotating}
\usepackage{caption}
\usepackage{subcaption}
\usepackage{epstopdf}
\usepackage{float}

\renewcommand{\vec}[1]{\mathbf{#1}}
\newcommand{\svec}[1]{\boldsymbol{#1}}
\newcommand{\Yv}{\vec{Y}}

\newcommand{\yv}{\vec{y}}

\newcommand{\xv}{\vec{x}}

\newcommand{\betav}{\svec{\beta}}

\newcommand{\bvh}{\hat{\svec{\beta}}}

\newcommand{\thetav}{\svec{\theta}}

\newcommand{\phiv}{\svec{\phi}}
\newcommand{\nuv}{\svec{\nu}}

\begin{document}
\begin{center}{\Large Predictive Likelihood for Coherent Forecasting of Count Time Series}\end{center}
\vspace{.5em}

\begin{center}{{\Large } $^a$Siuli Mukhopadhyay{\footnote {Corresponding author. Email: siuli@math.iitb.ac.in}, $^b$V. Satish}\\
{\it $^a$Department of Mathematics, $^b$ Department of Electrical Engineering, Indian Institute of Technology Bombay, }\\ {\it Mumbai 400 076, India}}\end{center}
\vspace{.25em}

\date{}


\hrule

\vspace{.5em}

\noindent {\bf Abstract}

A new forecasting method based on the concept of the profile predictive likelihood function is proposed for discrete valued processes. In particular, generalized auto regressive and moving average (GARMA) models for Poisson distributed data are explored in details. Highest density regions are used to construct forecasting regions. The proposed forecast estimates and regions are coherent. Large sample results are derived for the forecasting distribution. Numerical studies using simulations and a real data set are used to establish the performance of the proposed forecasting method. Robustness of the proposed method to possible misspecifications in the model is also studied. 
\vspace{.5em}

\vspace{.5em}

\vspace{.5em}
\section{Introduction}

In contrast to time series for Gaussian responses, where numerous forecasting methods are available, literature on forecasting for count type time series is still very sparse. However, time series data in form of counts is frequently measured in various fields like finance, insurance, biomedical, public health. 
As an example, consider a disease surveillance study, where health officials record the number of disease cases over a certain time period to understand the  disease trajectory. The main interest in such surveillance studies is to forecast the disease counts in the future, so that public health-care providers are able to respond to disease outbreaks on time, thereby reducing the  disease impact and saving economic resources (\cite{myers}). However, forecasting disease counts in these situations is complex, due to the fact that the required forecasts have to be consistent with the non-negative and integer valued sample space of such count time series. The usage of the estimated mean at a future time point which is a non integer, as a suitable point estimate for the future count (\cite{10.2307/30042088}, \cite{doi:10.1080/23737484.2016.1190307}, \cite{jalal})  as practiced in usual ARIMA forecasting techniques gives rise to an  incoherent forecast (\cite{FREELAND2004427}).

In this article, we present a forecasting method based on the profile predictive likelihood function for count time series data. The proposed forecast estimates and regions are coherent. The class of models we consider  is the observation driven models (\cite{10.2307/2336303}, \cite{Li1994}, \cite{10.2307/30045208}, \cite{Fokianos2004} and \cite{kedembook}), where the conditional distribution of counts given past information belongs to the exponential family. As done in the available statistical literature (\cite{Agresti1990}, \cite{Bishop1975}, \cite{Haberman1974}, \cite{Cameron1998}, \cite{Winkelmann2000}), we model the  counts with a Poisson type distribution. These Poisson time series models  allow inclusion of  both autoregressive and moving average terms along with trend, seasonality and dependence on several covariates. To the best of our knowledge this is the first coherent forecasting method proposed for count time series modeled by such observation driven models. 


Mainly two approaches have been used for modelling non Gaussian time series, parameter driven and observation driven models (\cite{cox}). These two types of models differ in the way they account for the autocorrelation in the data. In observation driven models, the
correlation between  observations measured at successive time points is modeled through a
function of past responses. However, parameter driven models use an unobserved latent process to 
 account for the  correlation. Conditioning on the
latent process, the observations are assumed to be evolve independently of the past
history of the observation process. Parameter driven models for non-Gaussian  time series were first considered by \cite{west}. Later, these models were also investigated by \cite{fahrmeir}, \cite{fahrwagen}, \cite{durbin1997,durbin2000}. 
To compute the posterior distributions for the parameters of these models  Markov chain Monte Carlo (MCMC) methods are frequently used (some references are \cite{durbin1997}, \cite{sp} and \cite{gamer}). However, often these MCMC algorithms fail to converge resulting in poor inference and predictions. In more recent times, the particle filter algorithm has been used as an alternative to the MCMC method in parameter driven models (\cite{dk}).  As compared to parameter driven models, the computational burden for parameter estimation   is much less in observation-driven models (\cite{10.2307/2291149},
1995; \cite{durbin2000}, \cite{10.2307/30042088}). 

In this article our focus is on  observation driven models, the conditional distribution for each observation given  past information on responses and past and possibly present covariates  is described by a generalized linear model (GLM) distribution. Partial likelihood theory combined with GLMs  is used for  estimation. This type of a flexible modeling framework  include namely, regression models for count time series proposed by \cite{10.2307/2336303}, GLM time series models  by \cite{Li1994} and \cite{Fokianos2004}, generalized linear autoregression models (GLMs) of \cite{shephard}, GLARMA models of \cite{10.2307/30042088} and GARMA models of \cite{10.2307/30045208}. However, in all these papers we may note that the main focus is on parameter estimation and in-sample prediction with almost no attention being given to the forecasting component involved. However, out-of-sample prediction based on past behaviour is an important and necessary component of time series studies. Consider a financial data based on number of transactions in stocks, the main interest in such studies  is to use the history of stock transactions and information on other external covariates influencing the stocks, to predict the number of future transactions and reap large profits (\cite{GRANGER19923}). In worldwide development of early warning systems for vector borne diseases like dengue, the main aim is to  forecast the future disease counts by
considering the effect of past disease outcomes and also different covariates like climate conditions, social ecological conditions
on disease counts (\cite{Shi} and \cite{Lee}). In both count time series mentioned above, number of stock transactions and disease cases, non-negative and integer valued estimates and intervals for future counts are required.



We use the profile predictive likelihood function of \cite{10.2307/4615723} to develop coherent forecasts for count time series. See \cite{bjornstad1990} and the references therein for a detailed review of some of the prediction functions used in the statistical literature. Forecasting using predictive likelihood has been used before for Gaussian models with an autoregressive structure of order one (\cite{10.2307/4616194}). However, to the best of our knowledge, it has not been used for out-of-sample forecasts in any non-Gaussian time series models.

Other forecasting techniques for count time series data modeled using the Poisson auto regressive model (PAR) and the integer auto regressive (INAR) class of models, have been proposed  by namely, \cite{FREELAND2004427}, \cite{MCCABE2005315}, \cite{10.2307/41057433} and \cite{doi:10.1111/stan.12083}.  However, the PAR and the INAR models are  structurally different from the observation driven GLM time series models as discussed above and usually do not accommodate covariates (\cite{mckenzie}). 

The original contributions of this article include (i)  coherent and consistent point forecasting method for count time series, (ii)  coherent forecasting regions based on highest density regions, (iii) use of the  the predictive likelihood concept for predicting dependent non Gaussian time series data. The performance of the proposed method is illustrated using several simulated examples and a real data set from a polio surveillance study. The effects of model misspecification on the proposed method and large sample results for the  forecasting estimator are also discussed. 

The rest of the paper is organized as follows. GLM for time series and particularly Poisson time series models are discussed in Section 2. This is followed by details on predictive likelihood (PL) function  and  PL for Poisson time series models in Section 3.   The steps of the forecasting algorithm, forecasting intervals based on highest density regions  and asymptotic properties of the forecast estimator are also discussed in this section. Section 4 gives the details of the simulation results, real data prediction results, and the study of the robustness of the proposed method. Concluding remarks are given in Section 5.


 \section{GLMs for time series}
The count at time $t$, $Y_t$ is assumed to depend on the past counts and past and possibly present covariate values. 
We assume that the conditional density of  each $y_t$ for $t=1,2,\ldots,n$, given the past information, $ H_{t} =\{x_t,\ldots,x_1,y_{t-1},\ldots,$ $y_1,\mu_{t-1},\ldots,\mu_1\}$, is given by,
\begin{equation}\label{A1}
 	f(y_{t}|H_{t})=exp\Big\{{\frac{y_{t}\gamma_{t}-b(\gamma_{t})}{\psi}+c(y_{t},\psi)}\Big\},
 	\end{equation}
 	where $ \gamma_{t} $ is the canonical  parameter, $\psi$ is the additional scale or dispersion parameter, while $ b(\cdot) $ and $ c(\cdot) $ are some specific functions corresponding to the type of exponential family considered. The canonical parameter $\gamma_t$ is a function of the conditional mean of $Y_t|H_t$, i.e., $\mu_{t}=b'(\gamma_{t})$. The conditional $Var(Y_t|H_t)$ is of the form $\sigma^{2}_{t}=\psi b''(\gamma_{t}) $. 
	The mean $\mu_t$ is related to the linear predictor $\eta_t$ by $\mu_t=h(\eta_t)$, where $h$ is a  one-to-one sufficiently smooth function. The inverse of $h$, denoted by $g$ is called the link function. To model the dependence of the means at time $t$, $\mu_t$, on $H_t$ the linear predictor is assumed to be,
	\begin{equation}\label{A2}
 	\eta_{t}=\xv'_{t}\betav+\sum_{j=1}^{p}\phi_{j}\mathcal{A}(y_{t-j},\xv_{t-j},\betav)+\sum_{j=1}^{q}\theta_{j}\mathcal{M}(y_{t-j},\mu_{t-j}),
 	\end{equation}
	where $\xv_t$ represents the covariate vector, $ \mathcal{A} $ and $ \mathcal{M} $ represent the autoregressive (AR) and moving average (MA) components, respectively. Some forms for the linear predictor under the general model setup proposed in the statistical literature are:\\
	Model form 1: \cite{10.2307/2336303}
	\begin{equation}
	\eta_t=\xv'_t\betav+\sum_{i=1}^{p}\phi_i [g(y_{t-i})-\xv'_{t-i}\betav], 
	\end{equation}
	Model form 2: \cite{Li1994} \begin{equation}
	\eta_t=\xv'_t\betav+\sum_{i=1}^{p}\phi_i y_{t-i}+\sum_{j=1}^{q}\theta_j\mu_{t-j},
	\end{equation}
	Model form 3: \cite{Fokianos2004} \begin{equation}
	\eta_t=\xv'_t\betav+\sum_{i=1}^{p}\phi_i g_i(y_{t-i})+\sum_{j=1}^{q}\theta_j d_j(\mu_{t-j}),
	\end{equation}
Model form 4: GARMA $(p,q)$ \cite{10.2307/30045208} \begin{equation}
	\eta_t=\xv'_{t}\betav+\sum_{j=1}^{p}\phi_j \{g(y_{t-j})-\xv'_{t-j}\betav\}+\sum_{j=1}^{q}\theta_j \{g(y_{t-j})-\eta_{t-j}\},\label{eta3}
\end{equation}
where $\xv_t$ is the vector of covariates at time $t$, $\betav$ are the usual regression parameters, $\phiv=(\phi_1,\ldots,\phi_p)'$ are the $p$ autoregressive parameters and $\thetav=(\theta_1,\ldots,\theta_q)'$ correspond to the $q$ moving average components, $g$, $g_i$ and $d_j$ are known functions.

Looking at the above model forms  in equations (3-6) we see that they are very similar to each other. We choose  the GARMA $(p,q)$ model proposed by \cite{10.2307/30045208} given in equation (\ref{eta3})  for representing the linear predictor. The GARMA models allows both autoregressive and moving average terms to be included in the linear predictor along with covariates.
In this article we model the conditional distribution of the count data using Poisson GARMA models, which we define next. 
	
\subsubsection{Poisson GARMA models}
The conditional density of $y_t$ given the past information is a Poisson distribution with mean parameter $\lambda_t$,
\begin{equation}
f(y_t|H_t)=\frac{\exp({-\lambda_t})\lambda_t^{y_t}}{y_t!},\,\lambda_t>0;\,y_t=0,1,2,\ldots;\,t=1,2,\ldots,n.
\end{equation}
Comparing with equation (\ref{A1}), we get $\gamma_t=\log(\lambda_t)$, $b(\gamma_t)=\lambda_t$, $\psi=1$ and $c(y_t,\psi)=-\log(y_t!)$. The conditional mean and variance of $Y_t$ given $H_t$ is $\lambda_t$. Using the GARMA $(p,q)$ model from equation (\ref{eta3}) and the canonical link function which is log for the Poisson family we write,
\begin{equation}log(\lambda_t)=\eta_t=\xv'_{t}\betav+\sum_{j=1}^{p}\phi_j \{g(y_{t-j})-\xv'_{t-j}\betav\}+\sum_{j=1}^{q}\theta_j \{g(y_{t-j})-\eta_{t-j}\},\label{3}\end{equation}
if the function $g$ is taken to be the log function, then for zero counts we define $y^{*}_{t-j}=\max(y_{t-j},c)$ where $0<c<1$ and 
\begin{equation*}log(\lambda_t)=\eta_t=\xv'_{t}\betav+\sum_{j=1}^{p}\phi_j \{\log(y^{*}_{t-j})-\xv'_{t-j}\betav\}+\sum_{j=1}^{q}\theta_j \{\log(y^{*}_{t-j}/\lambda_{t-j})\}.\label{3}\end{equation*}

\subsubsection{Estimation of Model Parameters}
As in a standard GLM, the maximum likelihood estimate (MLE) of  $\nuv=({\betav},{\phi}_1,\ldots,{\phi}_p,{\theta}_1,\ldots,{\theta}_q)'$, are obtained by using an iterated weighted least squares algorithm. Since here the  covariates $\xv_t$ are stochastic in nature, the partial likelihood function ($\prod_{t=1}^{n}f(y_t|H_t)$) instead of the entire likelihood function is maximized. 
An estimate of the conditional mean is given by, $\hat{\lambda}_t=\exp(\hat{\eta}_t)$, where \begin{equation*}\label{A2}
 	\hat{\eta}_{t}=\xv'_{t}\bvh+\sum_{j=1}^{p}\hat{\phi}_j \{\log(y^{*}_{t-j})-\xv'_{t-j}\bvh\}+\sum_{j=1}^{q}\hat{\theta}_j \{\log(y^{*}_{t-j}/\hat{\lambda}_{t-j})\},
 	\end{equation*}
$\hat{\nuv}$ is the MLE of $\nuv$. For the Poisson GARMA model, the conditional information matrix of the parameter estimates is given by $I=\lim_{n\rightarrow\infty}\frac{1}{n}\left[\sum_{t=1}^{n}\lambda_t \left(\frac{\delta\eta_t}{\delta\nuv}\right)\left(\frac{\delta\eta_t}{\delta\nuv}\right)'\right]^{-1}$. 
 
%
%
%

\section{Predictive Likelihood}

Prediction of the value of an observation at a future time point is a fundamental problem in statistics, especially in the time series context. To be more precise, suppose we have observations at time points $t_1,\ldots,t_n$, denoted by $\yv=(y_1,\ldots,y_n)'$ and we would like to predict or forecast (out-of-sample) the observation at time point $t_{n+m}, m=1,2,\ldots$. We assume that $(\Yv,Y_{n+m})$ has a joint probability density $f_\omega(\yv,y_{n+m})$ indexed by the unknown parameter $\omega$. Note in this prediction problem we have two unknown quantities, $y_{n+m}$, the unobserved value of $Y_{n+m}$ and the parameter, $\omega$. However, our primary aim is to find $Y_{n+m}$. So we treat $\omega$ as a  nuisance parameter. \cite{10.2307/4615723} proposed several predictive functions for finding the future value $y_{n+m}$ given the observed sample $\yv$. 
   In this article we choose the  prediction function, say $LL_p$, from the likelihood viewpoint,
 \begin{equation*}
 {LL}_p(y_{n+m}|\yv)=\sup_\omega f_\omega(\yv,y_{n+m})=f(\yv,y_{n+m}|\hat{\omega}_{y_{n+m}}),
 \end{equation*}
 for inferring about the unknown $y_{n+m}$. In the above equation, the maximum likelihood estimate (MLE) of ${\omega}$ denoted by $\hat{\omega}_{y_{n+m}}$ is computed  using the $n$ available observations $\yv$ and  the unobserved future count $y_{n+m}$. 
 
 Different ways of estimating the nuisance parameters lead to different predictive likelihoods. 
 For our problem we use the MLE method to estimate and eliminate the dependence on the nuisance parameter $\omega$.  This particular form of the prediction function is known as the profile predictive likelihood (PL) function. \cite{10.2307/2335744}, \cite{10.2307/4615564}, \cite{hinkley1979} and \cite{10.2307/2287730}, suggested an alternative method for the elimination of the unknown parameter $\omega$ by using the principle of sufficiency. However, finding closed form  sufficient statistic for the parameters is impossible given the complicated  framework of the GARMA models  and the correlated nature of the responses.

 \subsection{PL for the Poisson Time Series}
 
A sample of size $n$ is observed from the Poisson GARMA $(p,q)$ model and denoted  by $\yv=(y_1,\ldots,y_n)'$. We start with the one step at a time prediction (out-of-sample forecasts) of counts.  In the context of the Poisson GARMA time series model, the predictive likelihood function is 
 \begin{equation}
 {LL}_p(y_{n+1}|H_{n+1})=\sup_{\nuv} f_{\nuv}(y_{n+1},H_{n+1})=\sup_{\nuv} \prod_{t=1}^{n+1}f_{\nuv}(y_t|H_t)=\frac{\exp(-\{\sum_{t=1}^{n+1}\hat{\lambda}_t\})\prod_{t=1}^{n+1}\hat{\lambda}_t^{y_t}}{\prod_{t=1}^{n+1}y_t!},
 \end{equation}
 where \begin{equation*}\hat{\lambda}_t=\exp(\hat{\eta}_t)=\exp(\xv'_{t}\bvh+\sum_{j=1}^{p}\hat{\phi}_j \{\log(y^{*}_{t-j})-\xv'_{t-j}\bvh\}+\sum_{j=1}^{q}\hat{\theta}_j \{\log(y^{*}_{t-j})-\hat{\eta}_{t-j}\}).\label{3}\end{equation*} is the MLE of $\lambda_t$ using the observed sample $\yv$ and the future count $y_{n+1}$. Here, $H_{n+1}=\{x_{n+1},\ldots,x_1,y_{n},\ldots,$ $y_1,\lambda_{n},\ldots,\lambda_1\}$, is the available past information at time $n+1$. The estimated density of $Y_{n+1}$ given the observed time series is then,
 \begin{equation}
 \hat{p}(y_{n+1})=k(H_{n+1}){LL}_p(y_{n+1}|H_{n+1}),\label{densityest}
 \end{equation}
 where $k(\cdot)$ is the normalizing constant. 
 
 The predictor of  $y_{n+1}$ is chosen to be the value of $y_{n+1}$ which maximizes $\hat{p}(y_{n+1})$, and denoted by $\hat{Y}_{n+1(PL)}$. The steps leading to the choice of $\hat{Y}_{n+1(PL)}$ are detailed below and later illustrated with an example:
	\begin{enumerate}
	\item At the start of the process we have the past information $H_{n+1}$. Our interest is to compute ${p}(y_{n+1})$, the actual density of $Y_{n+1}$ given the past information.  However the density depends on the parameters $\lambda_t$ which are unknown. So we first compute the MLEs of the parameters and then use equation (\ref{densityest}) to compute an estimate of ${p}(y_{n+1})$.  
		\item To compute the MLEs of $\lambda_{t},\,t=1,\ldots,n+1$ we need $H_{n+1}$ and $y_{n+1}$. We have not yet observed $y_{n+1}$. So we start with the first possible value i.e., $ y_{n+1} =0$ and estimate the $\lambda_t$'s and calculate $\hat{p}(y_{n+1}=0)$. 
		\item We next set $ y_{n+1} = 1$ and re-estimate the parameters and find the corresponding normalized predictive likelihood value  $\hat{p}(y_{n+1}=1)$.
		\item The above step is repeated  for $y_{n+1}=2,3,\ldots$ , each time re-estimating the parameters and computing $\hat{p}(y_{n+1}=i)$. If  $\hat{p}(y_{n+1}=i)<1\times 10^{-6}$ for any $i$, then we  ignore $y_{n+1}=i$. 
				\item At the end we have a set of $q$ (say) normalized predictive likelihoods of $y_{n+1}$, denoted by  $ \hat{p}(y_{n+1}=0), \hat{p}(y_{n+1}=1), \ldots, \hat{p}(y_{n+1}=q-1)$.

If $\hat{p}(y_{n+1}=d)=\max\{\hat{p}(y_{n+1}=0), \hat{p}(y_{n+1}=1), \ldots, \hat{p}(y_{n+1}=q-1)\}$, we  set $ \hat{Y}_{n+1(PL)}  = d$.
\end{enumerate}

{\bf Example}: We explain the above steps of the algorithm with a simple example. A sample of size $100$ is drawn from a Poisson GARMA $(0,2)$ model, denoted  by $\yv=(y_1,\ldots,y_{100})'$ and the model equation is 
\begin{eqnarray*}\label{}         
	&\log(\lambda_{t})& =0.2+0.01 t+0.4 \cos(2\pi t/12)+0.5 \sin(2\pi t/12)+0.5\cos(2\pi t/6)\\&&+0.5\sin(2\pi t/6)-0.5\log(y^{*}_{t-1}/\lambda_{t-1})+0.6\log(y^{*}_{t-2}/\lambda_{t-2}),
\end{eqnarray*}      for  $y_{t-j}^{*} = \max(y_{t-j}, 0.1),\,j=1,2$. 
We apply the PL procedure as given above for predicting $Y_{101}$. Table \ref{example} shows the $q$ (19) possible values of $y_{101}$ and the corresponding density value. Note $\hat{p}(y_{101})<1\times 10^{-6}$ for $y_{101}>19$.
\begin{table}[h!]
	\centering
	\begin{tabular}{|l|l|l|l|l|l|l|l|}
		\hline
		$y_{101}$ & $\hat{p}(y_{101})$ & $y_{101}$ & $\hat{p}(y_{101})$ & $y_{101}$ & $\hat{p}(y_{101})$ & $y_{101}$ & $\hat{p}(y_{101})$ \\ \hline
		0         & 0.715$\times 10^{-2}$    & 5         & 0.1754                   & 10        & 0.170$\times 10^{-1}$    & 15        & 0.139$\times 10^{-3}$    \\ \hline
		1         & 0.353$\times 10^{-1}$    & 6         & 0.1443                   & 11        & 0.765$\times 10^{-2}$    & 16        & 0.429$\times 10^{-4}$    \\ \hline
		2         & 0.873$\times 10^{-1}$    & 7         & 0.1018                   & 12        & 0.315$\times 10^{-2}$    & 17        & 0.124$\times 10^{-4}$    \\ \hline
		3         & 0.14378                  & 8         & 0.629$\times 10^{-1}$    & 13        & 0.119$\times 10^{-2}$    & 18        & 0.342$\times 10^{-5}$    \\ \hline
		4         & 0.1775                   & 9         & 0.345$\times 10^{-1}$    & 14        & 0.422$\times 10^{-3}$    & 19        & 0.89$\times 10^{-6}$     \\ \hline
	\end{tabular}
\caption{$y_{101}$ and corresponding $\hat{p}(y_{101})$ values}
\label{example}
\end{table}

The maximum value of $\hat{p}(y_{101})$ is 0.1775 corresponding to $y_{101}$  = 4. Thus, predicted value of $y_{101}$ is $ \hat{Y}_{101(PL)}  = 4$.
 

The PL function can also be extended for m-step predictions of Poisson GARMA models.
 The predictive likelihood function for $m$ step ahead forecasts  is,
 \begin{eqnarray}\label{mstep}
\nonumber {LL}_p(y_{n+m}|H_{n+m})&=&\sup_{\nuv} f_{\nuv}(H_{n+m},y_{n+m})=\sum_{y_{n+m-1}}\ldots\sum_{y_{n+1}}\sup_{\nuv} \prod_{t=1}^{n+m}f_{\nuv}(y_t|H_t)\\&=&\sum_{y_{n+m-1}}\ldots\sum_{y_{n+1}}\frac{\exp(-\sum_{t=1}^{n+m}\hat{\lambda}_t)\prod_{t=1}^{n+m}\hat{\lambda}_t^{y_t}}{\prod_{t=1}^{n+m}y_t!},
 \end{eqnarray}
 where the MLEs of $\lambda_t$ are found using the observed sample $\yv$ and the past information $H_{n+m}$. The value of $y_{n+m}$ which maximizes the normalized $\hat{p}(y_{n+m})$ is chosen as the predictor of $y_{n+m}$ and denoted by $\hat{Y}_{n+m(PL)}$. 
 
We note from the above forecasting equations that to predict a future count at time $n+m$, we need information on the counts, $y_t$ and also on the covariates $\xv_t$ at time points $n+1,\ldots,n+m-1$.  The covariates may be stochastic in nature, say in a disease data set we study the effect of time and also covariates like temperature and humidity on the disease counts. In these situations, separate time series models have to be fitted to the covariates data and used for obtaining forecasts of the covariates. These forecasted values of the covariates can then be used in the PL function to forecast $y_{n+m}$. 
 
 \subsection{Forecast Region}
In the last section we used the predictive likelihood to get a point forecast of $Y_{n+m}$ by using the data $Y_1,\ldots, Y_n$. Along with the point forecast it is of our interest also to identify a region in the sample space of $Y_{n+m}|H_{n+m}$ which in some sense will summarize the actual density $p(y_{n+m})$. 
 Such a forecasting region can be constructed in several ways. 
 
We may suggest that the $(1-\alpha)$ predictive interval  of $Y_{n+m}$ given by $[\hat{z}(\alpha/2),\hat{z}(1-\alpha/2)]$, where
 $\hat{z}(\alpha)$ is the $\alpha$ th quantile of $p(y)$, is one such region. 
However, since we are dealing with an asymmetric distribution here, it would be more prudent to use highest density regions (HDRs) instead of a forecasting region based on quantiles. 
These regions are more flexible than those based on quantiles and are able to address the issues of  asymmetry, skewness as well as multimodality in the forecast distributions.
 The $(1-\alpha)100\%$ HDR denoted by $R(f_\alpha)$ is the region in the sample space of $Y_{n+m}$  such that (\cite{10.2307/2684423})
\begin{equation*}
R(f_\alpha)=p(y_{m+n})\geq f_\alpha,
\end{equation*}
 where $f(\alpha)$ is the largest constant such that $Pr(Y_{n+m}\in R(f_\alpha))$ is at least $(1-\alpha)$. 
 In our computations of HDRs, we used the $\alpha$th sample quantile of $\hat{p}(y_{m+n})$ to estimate 
$\hat{f}_\alpha$ and then  obtained ${\hat{R}(\hat{f}_\alpha)}$ by choosing the values of $y_{n+m}$ for which $\hat{p}\geq \hat{f}_\alpha$.  

{\bf Note}: that the sample space of  $y_{n+m}$ is non negative and integer valued. Thus,  $y_{n+m}$ can take only the integer values in the computed HDRs giving rise to coherent forecasting regions.

\subsection{Some large sample results}

In this section we discuss the consistent properties and the asymptotic distribution of the forecast estimator. 

{\bf Proposition 1}: As $n\rightarrow\infty$, $\hat{p}(y_{n+m})\xrightarrow{P}p(y_{n+m})$. 

Proof: We use Theorem 2.3.5 from \cite{sen1994large} which states

If $ T_n \xrightarrow{P} T $ and if $g(t)$ is uniformly continuous, then $g(T_n)\xrightarrow{P}g(T)$.

For our problem, $T=\nuv$ and $T_n=\hat{\nuv}$. Also $g(T_n)=\hat{p}(y_{n+m})$ and $g(T)={p}(y_{n+m})$.  

The normalized predictive likelihood function $\hat{p}(y_{n+m})=k(H_{n+m}){LL}_p (y_{n+m}|H_{n+m})$ is uniformly continuous since 
\begin{equation*}
\frac{d\hat{p}(y)}{d\hat{\nuv}}=\sum_{y_{n+m-1}}\ldots\sum_{y_{n+1}}\frac{\exp(-\sum_{t=1}^{n+m}\hat{\lambda}_t)\prod_{t=1}^{n+m}\hat{\lambda}_t^{y_t}}{\prod_{t=1}^{n+m}y_t!}\sum_{t=1}^{n+m}\Big(\frac{y_{t}}{\hat{\lambda}_{t}} - 1\Big)\frac{d\hat{\lambda}_{t}}{d\hat{\nuv}}
\end{equation*}
exists and is bounded. 

Thus proved.

{\bf Proposition 2}: As $n\rightarrow\infty$, 
\begin{equation*}
\sqrt{n-m}[\hat{p}(y_{m+n})-p(y_{m+n})]\xrightarrow{d}N(0,\nabla p^{T}(y_{n+m})I^{-1}(\nuv)\nabla p(y_{n+m})).
\end{equation*}

Proof: We know that the ML estimator ($ \hat{\nuv} $) is asymptotically normal (\cite{10.2307/30045208}),
\begin{equation*}
\sqrt{n-m}[ \hat{\nuv} -\nuv]\xrightarrow{d}N(0, I^{-1}(\nuv)).
\end{equation*}
Using the  the first order approximation of a Taylor series,
\begin{equation*}
\hat{p}(y_{n+m}) \approx p(y_{n+m}) + \nabla p^{T}(y_{n+m}) (\hat{\nuv} - \nuv),
\end{equation*}
Rearranging the terms and multiplying both sides by $ \sqrt{n-m} $ we obtain,
\begin{equation*}
\sqrt{n-m}[\hat{p}(y_{n+m}) - p(y_{n+m})] = \nabla p^{T}(y_{n+m})\sqrt{n-m} (\hat{\nuv} - \nuv),
\end{equation*}
From Slutsky's theorem it follows,
\begin{equation*}
\sqrt{n-m}[\hat{p}(y_{n+m})-p(y_{n+m})]\xrightarrow{d}N(0,\nabla p^{T}(y_{n+m}) I^{-1}(\nuv)\nabla p(y_{n+m}))
\end{equation*}
This concludes the proof.

{\bf Note}: Though in Propostion 2 we show that the large sample distribution of the forecast estimate is a Gaussian distribution, this fact has not been used anywhere in the computations. We have used the Poisson mass function for all the computations.
  
 
 \section{Results}
 In this section we illustrate the proposed forecasting technique using simulation studies and a real data set based on polio counts. Robustness properties of the PL forecasts are also studied.
 
 \subsection{Simulations}
 The performance of the predictive likelihood in forecasting GARMA $(p,q)$ models is evaluated using simulation studies for various values of $p$ and $q$. Along with point prediction, we also compute  HDRs for the future counts. 
 
 Our simulation results mainly focus on one step at a time forecasts instead of $m$ step ahead forecasts. Note, for computing $m$ step ahead forecasts using equation (\ref{mstep})  complete enumeration of all the sums and products over all the possible projected paths from $n+1$ to $n+m-1$ is necessary which is impossible unless we truncate the sample space.  We discuss one simulated example for 2 step ahead forecasting based on a truncated sample space. 
 
 Two different GARMA models with the following means were used in the simulations:
\begin{itemize}
\item {\bf Model 1}: GARMA $(5,0)$
\begin{eqnarray*}\label{}         
	&\log(\lambda_{t})& = \text{x}_{t}^{'}\betav+\sum_{j=1}^{5}\phi_{j}\{\log(y^{*}_{t-j})-\text{x}_{t-j}^{'}\betav\},
\end{eqnarray*}
where  $\text{x}_{t} =  [\begin{matrix}1& t& \cos(2\pi t/12)& \sin(2\pi t/12)& \cos(2\pi t/6)& \sin(2\pi t/6) \end{matrix}]^{'}$,\\ $\betav = [\begin{matrix} 0.2&0.001 & 0.5& -0.5& 0.6& 0.7 \end{matrix}]^{'}$ and $\phi_{1} = 0.5 ,\phi_{2} = -0.6,\phi_{3} = 0.4,\phi_{4} = -0.6,\phi_{5} = 0.5$.

\item {\bf Model 2}: GARMA $(0,2)$ \begin{eqnarray*}\label{}         
&\log(\lambda_{t})& =0.2+0.01 t+0.4 \cos(2\pi t/12)+0.5 \sin(2\pi t/12)+0.5\cos(2\pi t/6)\\&&+0.5\sin(2\pi t/6)-0.5\log(y^{*}_{t-1}/\lambda_{t-1})+0.6\log(y^{*}_{t-2}/\lambda_{t-2}),
\end{eqnarray*}      
\end{itemize}
for  $y_{t-j}^{*} = \max(y_{t-j}, 0.1),\,j=1,2$. 
For both models, we chose three different values of $n=50,100,240$. The different values of $n$ enabled us to study the effect increasing $n$ on the profile likelihood predictions. Ten future counts (time horizon is 10 counts) were predicted one step at a time in each case. It is possible to increase the time horizon to more than 10, however those computations are not shown here.

We explain the simulation steps for a Poisson GARMA $(p,q)$ model in details below for a fixed value of $n$:
\begin{enumerate}
\item Using the chosen GARMA model and the fixed value of $n$ we simulated $N_0$ data sets. For all computations  $N_0$ is 1000. Increasing $N_0$ further did not change the results.
 \item The PL predictor of $y_{n+m}$ in the $i$th simulation  is denoted  by $\hat{Y}^{(i)}_{n+m(PL)}$ for $i=1,\ldots,N_0$. We also get $N_0$ HDRs for $y_{n+m}$.  The minimum and maximum values of $y_{n+m}$ in the $i$th HDR is denoted as $HDR^{(i)}(min)$ and  $HDR^{(i)}(max)$, respectively.
 \item The  $\text{median}\{\hat{Y}^{(i)}_{n+m(PL)},\,i=1(1)N_0\}$ is selected as the predicted value of $y_{n+m}$. 
Selecting the median instead of the mean ensures a integer future count. 
\item The  median of  $\{HDR^{(i)}(min),\,i=1(1)N_0\}$ is selected as the lowest value of $y_{n+m}$ in the HDR and denoted by $Ly_{n+m}$, while  
median of  $\{HDR^{(i)}(max),\,i=1(1)N_0\}$ is chosen as the highest value and denoted by $Uy_{n+m}$. The HDR is then defined as  $\{Ly_{n+m},Uy_{n+m}\}$. Remember $y_{n+m}$ is discrete valued. For clarity, if $Ly_{n+m}=1$ and $Uy_{n+m}=3$, then the HDR is $\{1,2,3\}$.
Note, all the HDRs computed displayed unimodal behavior.

\end{enumerate}

 The true and predicted values of the ten future counts and the 50 and 75$\%$ HDRs  are reported in Figures 1-2 for simulation models 1 and 2, respectively. Table \ref{RMSE} reports the RMSE values for both simulation models. From the figures and the table it is noted that  point prediction  using the profile likelihood method improves as $n$ increases. This happens since the MLE, $\hat{\nuv}$,  converges to the true $\nuv$ as $n$ increases, which in turn causes the normalized ${LL}_p$ to converge to the actual pdf of $Y_{n+m}$ (see Proposition 1). Also we note that the HDRs successfully capture all the true counts in both simulation models for  $n=100,240$. For $n=50$, the $50$ and $75\%$ HDRs are unable to capture $y_{55}$, however the $95\%$ HDRs (not shown) are able to capture $y_{55}$. 
 
 \begin{figure}
	\centering
	\begin{subfigure}[b]{0.495\textwidth}
		\centering
		\includegraphics[scale=0.13]{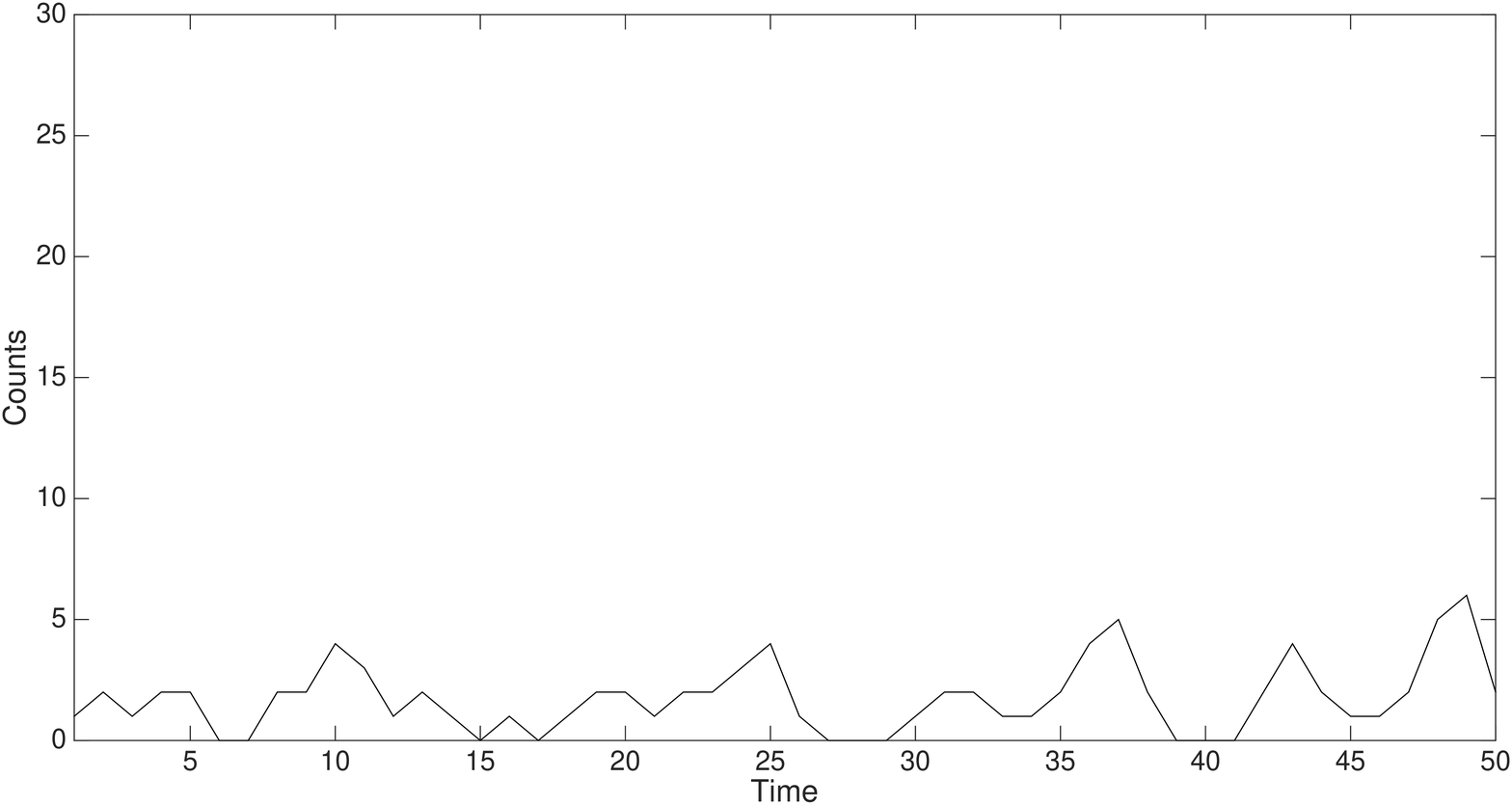}
		\caption{Simulated data, $y_1,\ldots,y_{50}$}
		\label{}
	\end{subfigure}
	\begin{subfigure}[b]{0.495\textwidth}
		\centering
		\includegraphics[scale=0.13]{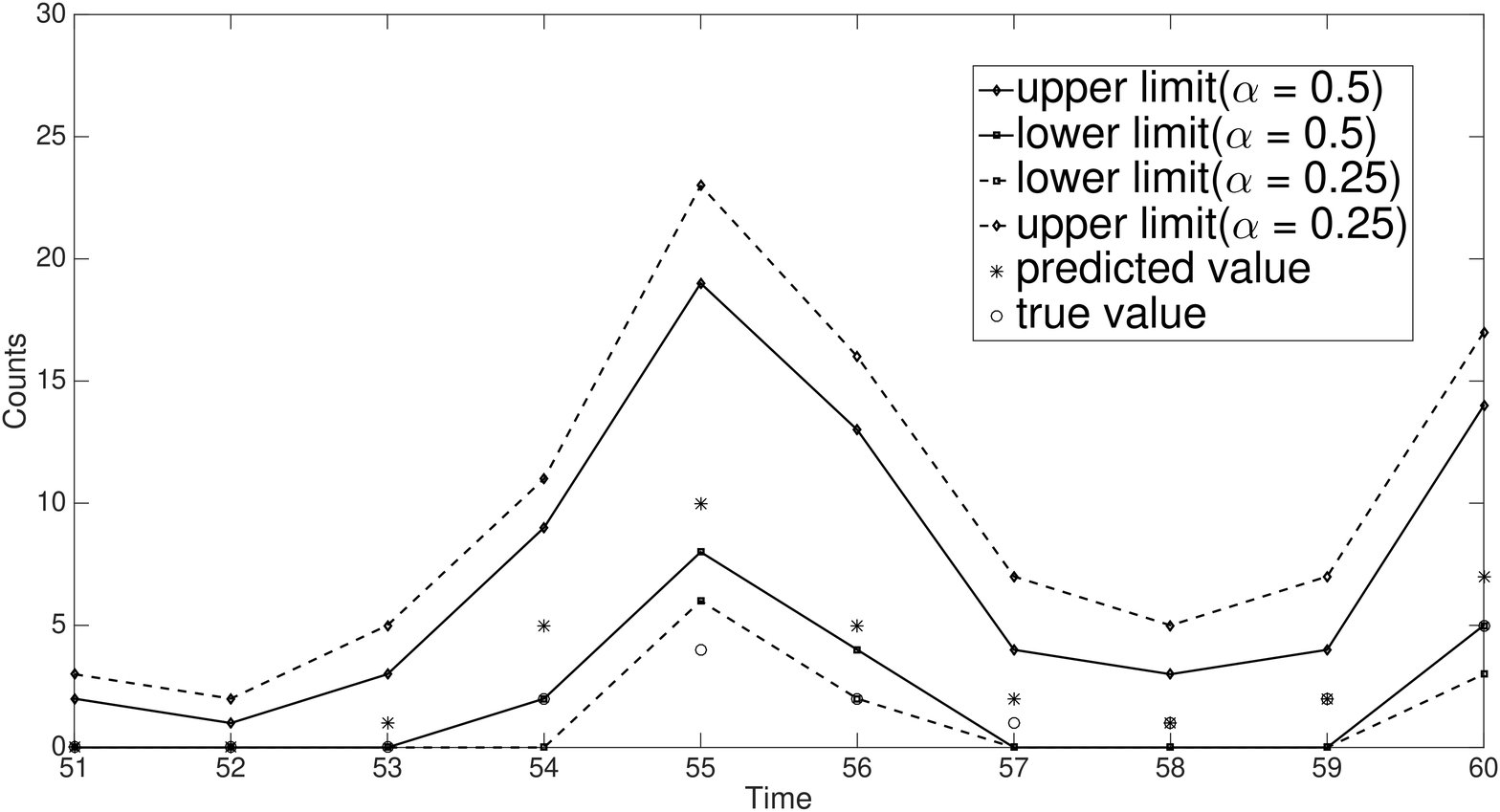}
		\caption{Forecasting results for $y_{51},\ldots,y_{60}$}
		\label{}
	\end{subfigure}
\quad
\vfill
	\begin{subfigure}[b]{0.495\textwidth}
		\centering
		\includegraphics[scale=0.13]{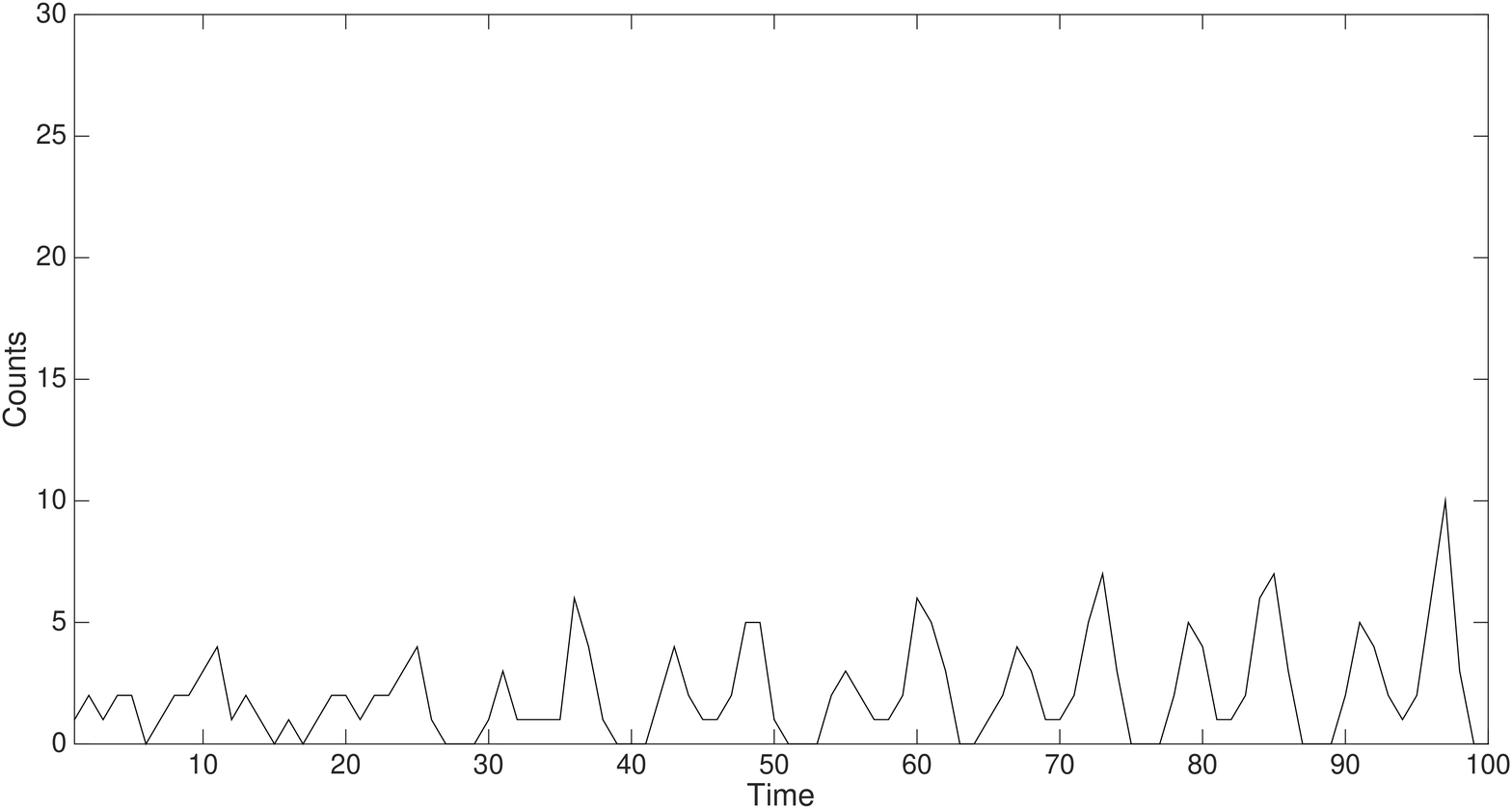}
		\caption{Simulated data, $y_1,\ldots,y_{100}$}
		\label{}
	\end{subfigure}
	\begin{subfigure}[b]{0.495\textwidth}
		\centering
		\includegraphics[scale=0.13]{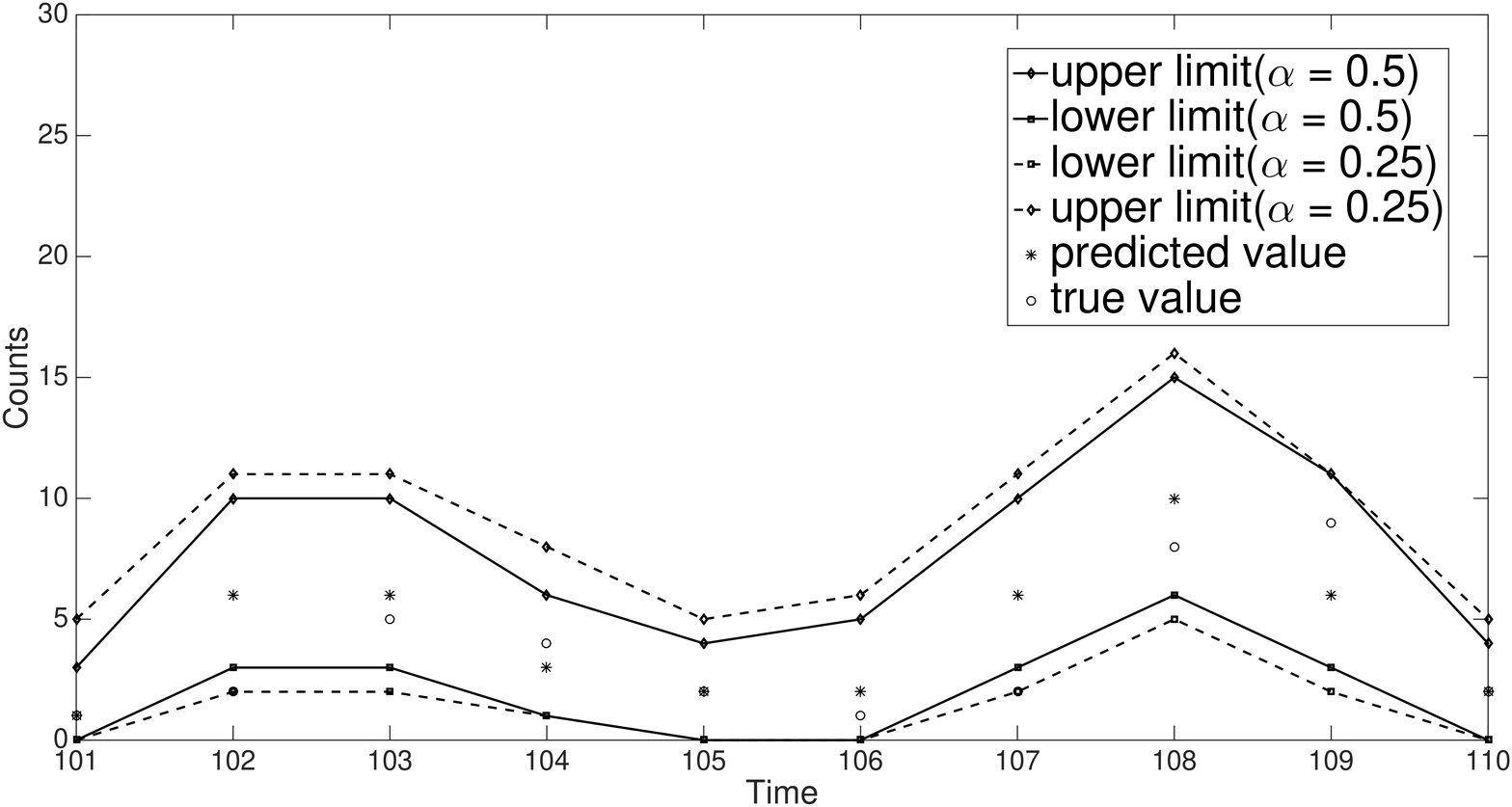}
		\caption{Forecasting results for $y_{101},\ldots,y_{110}$}
		\label{}
	\end{subfigure}
\quad
\vfill
	\begin{subfigure}[b]{0.495\textwidth}
		\centering
		\includegraphics[scale=0.13]{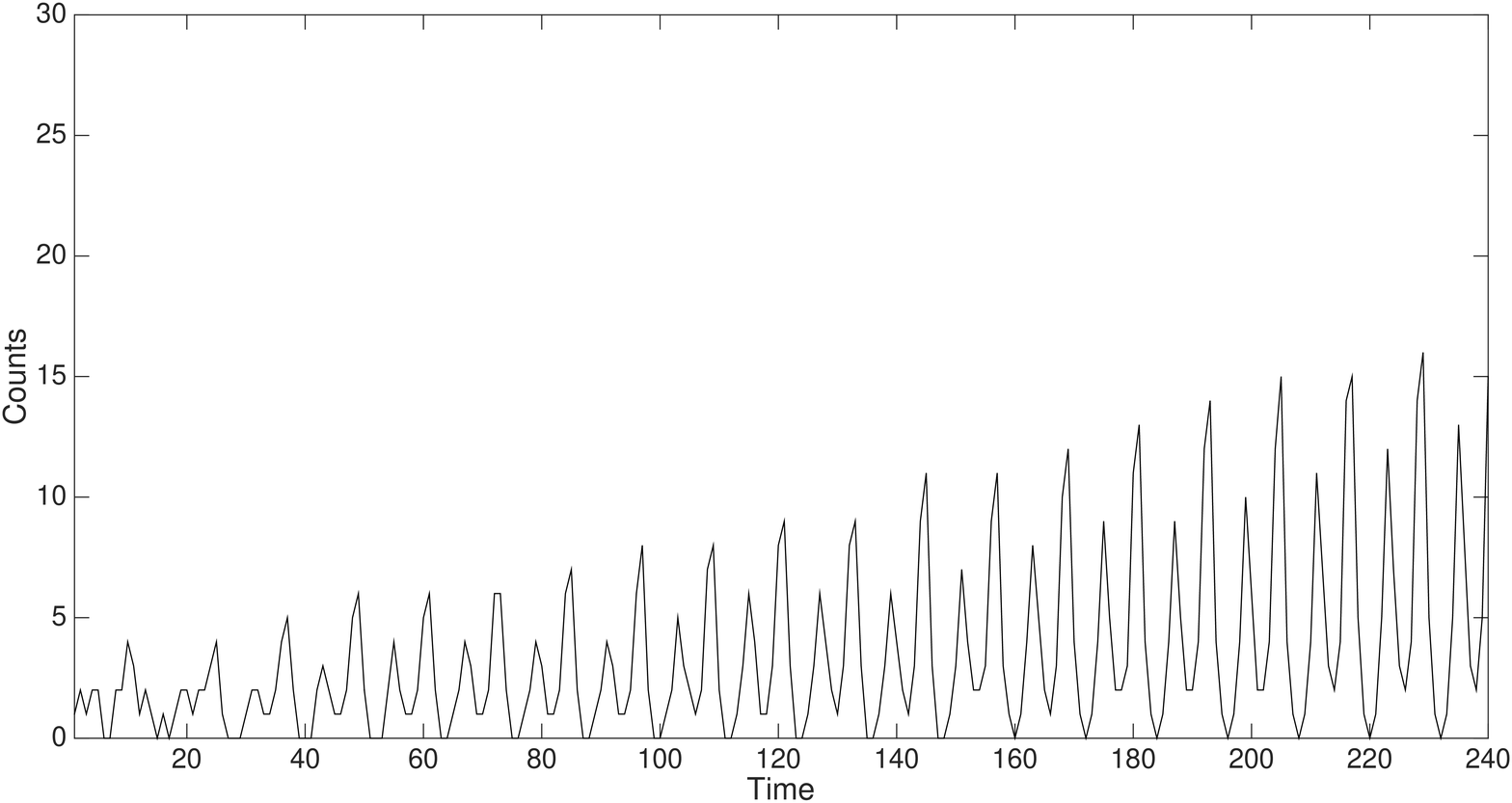}
		\caption{Simulated data, $y_1,\ldots,y_{240}$}
		\label{}
	\end{subfigure}
	\begin{subfigure}[b]{0.495\textwidth}
		\centering
		\includegraphics[scale=0.13]{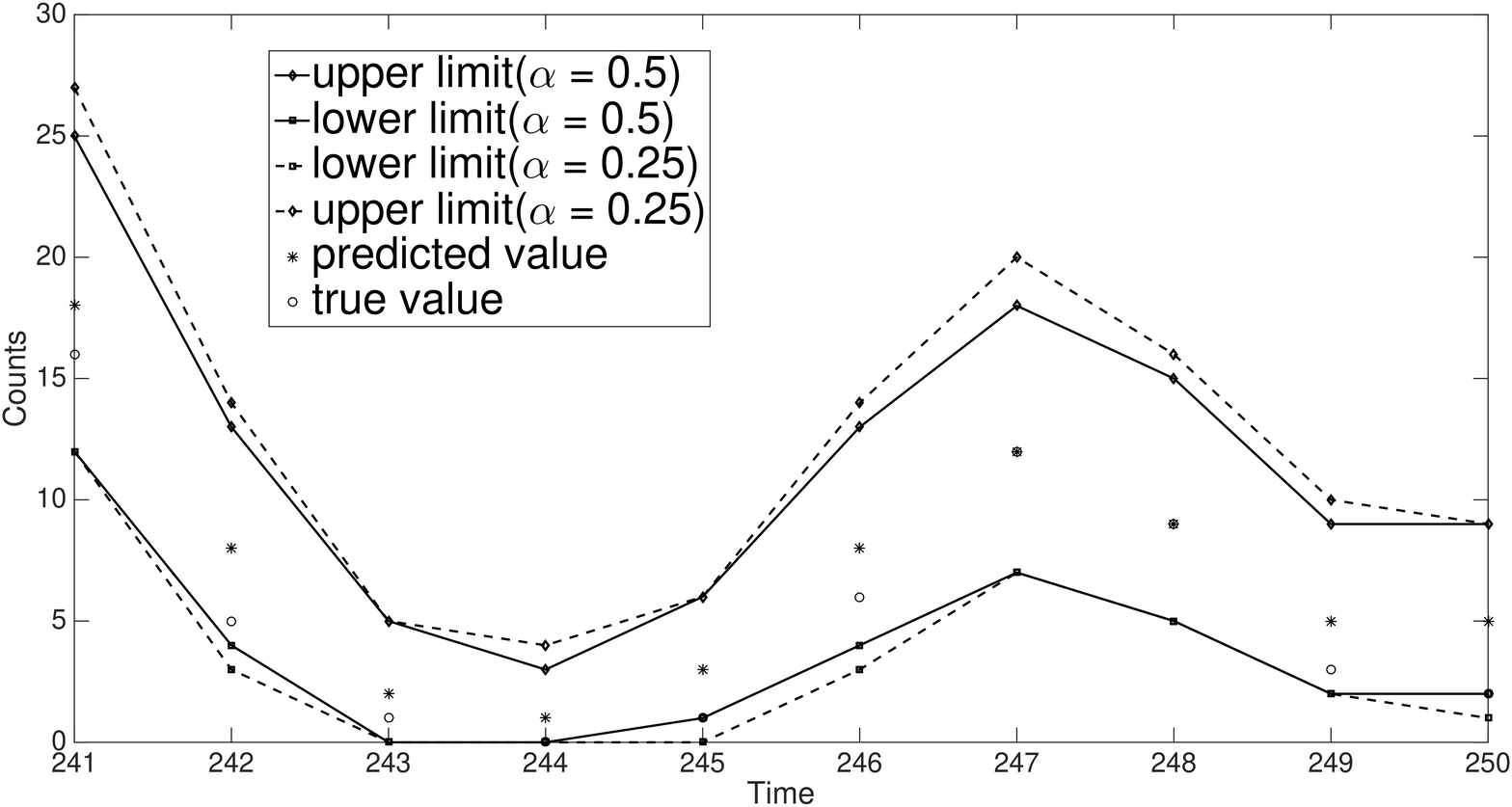}
		\caption{Forecasting results for $y_{241},\ldots,y_{250}$}
		\label{}
	\end{subfigure}
\caption{Simulation and Forecasting results using the PL function for the GARMA $(5, 0)$ model. Note the HDRs of $y_{n+m}$ though shown as continuous intervals contain only integer values.}
\label{}
\end{figure}

\begin{figure}
	\centering
	\begin{subfigure}[b]{0.495\textwidth}
		\centering
		\includegraphics[scale=0.13]{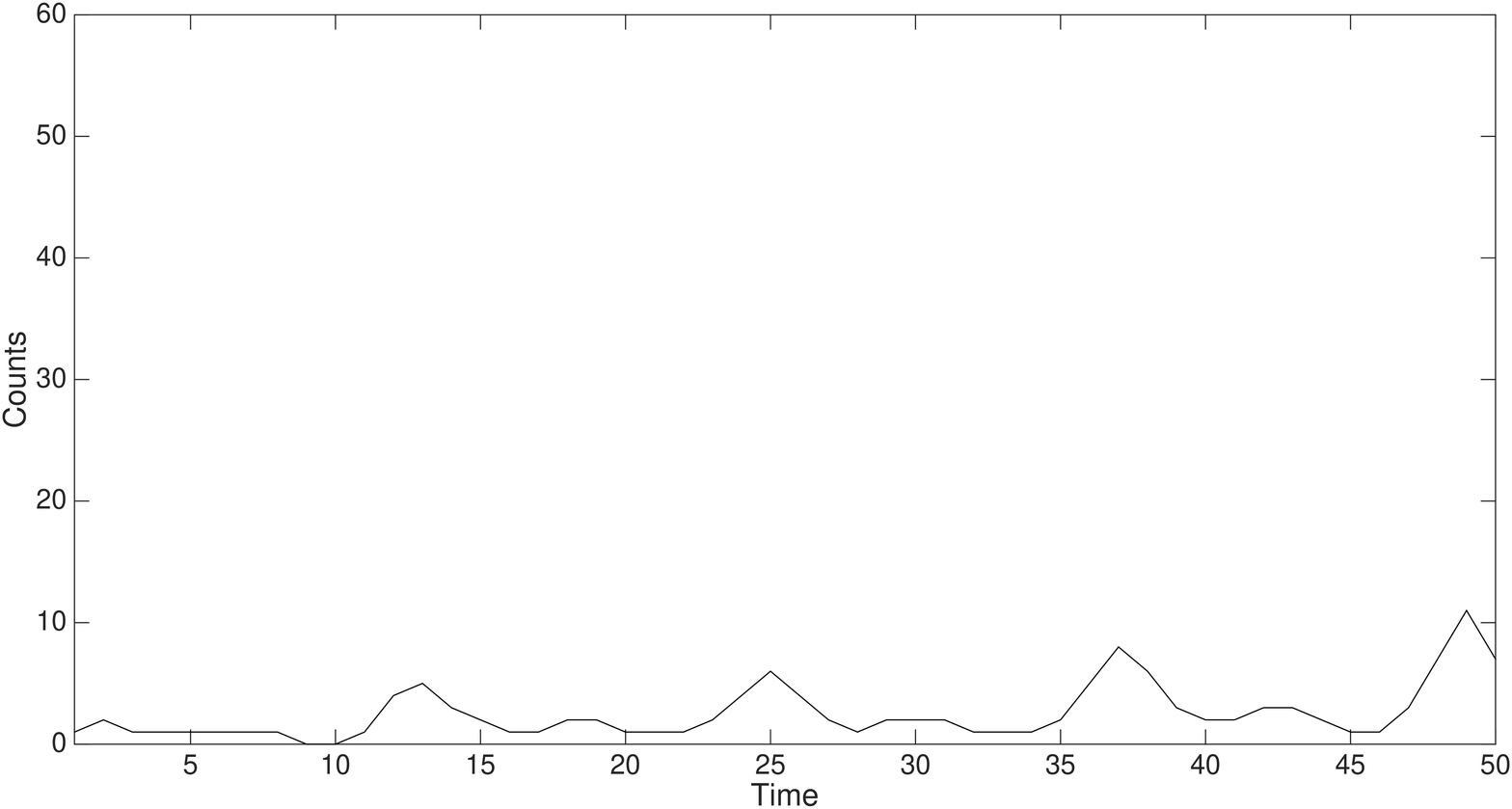}
		\caption{Simulated data, $y_1,\ldots,y_{50}$}
		\label{}
	\end{subfigure}
	\begin{subfigure}[b]{0.495\textwidth}
		\centering
		\includegraphics[scale=0.13]{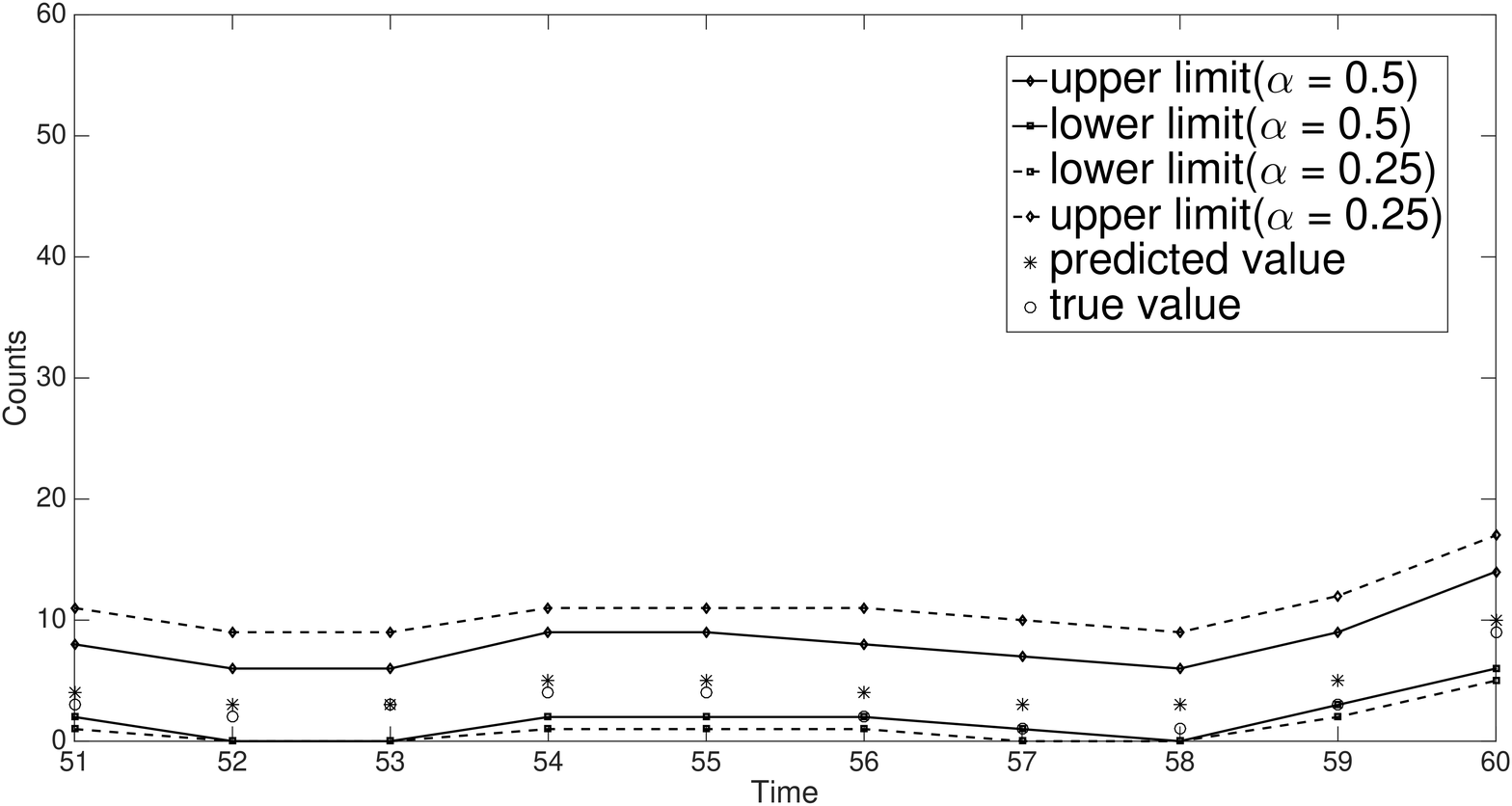}
		\caption{Forecasting results for $y_{51},\ldots,y_{60}$}
		\label{}
	\end{subfigure}
\quad
\vfill
	\begin{subfigure}[b]{0.495\textwidth}
		\centering
		\includegraphics[scale=0.13]{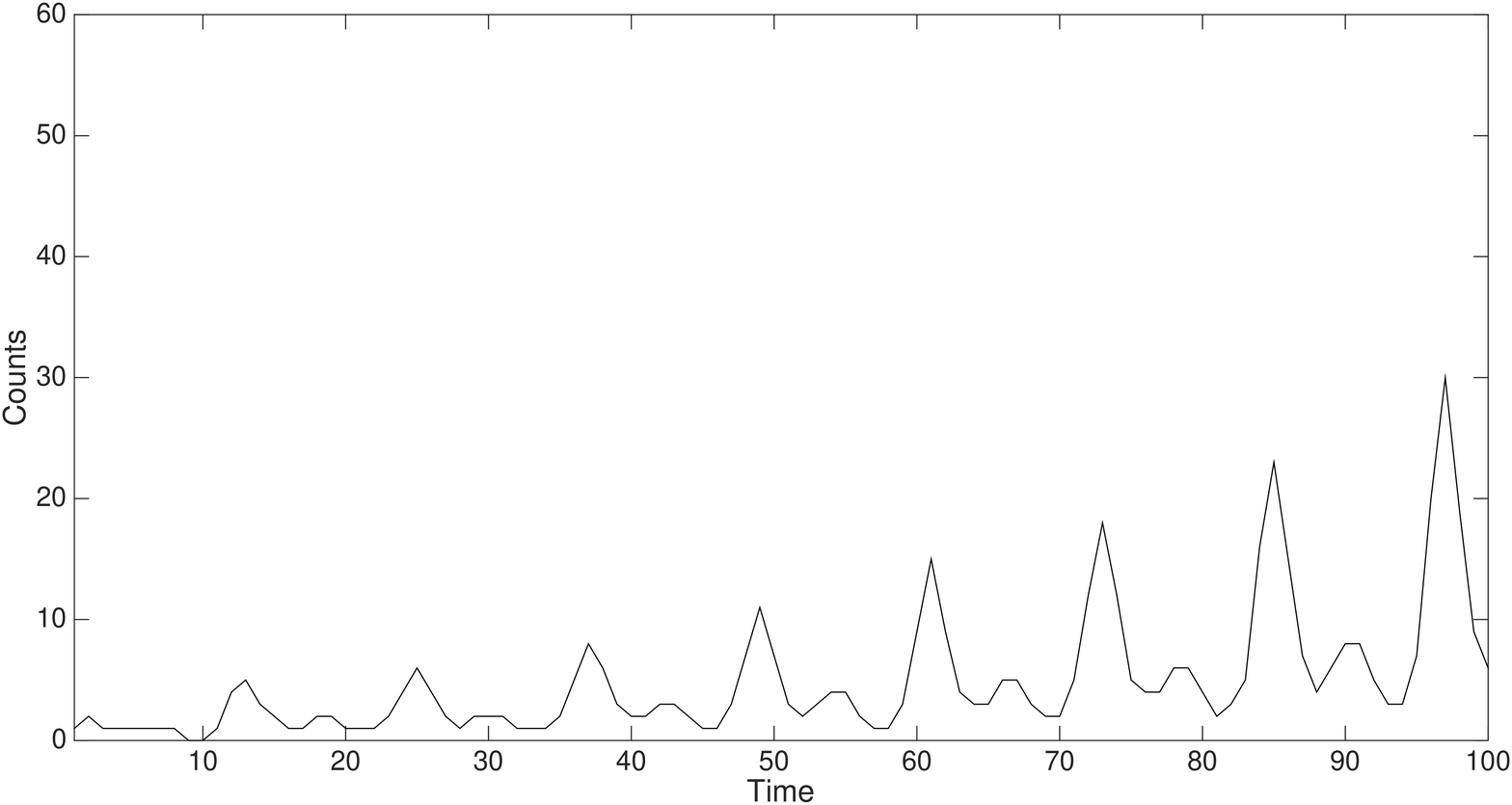}
		\caption{Simulated data, $y_1,\ldots,y_{100}$}
		\label{}
	\end{subfigure}
	\begin{subfigure}[b]{0.495\textwidth}
		\centering
		\includegraphics[scale=0.13]{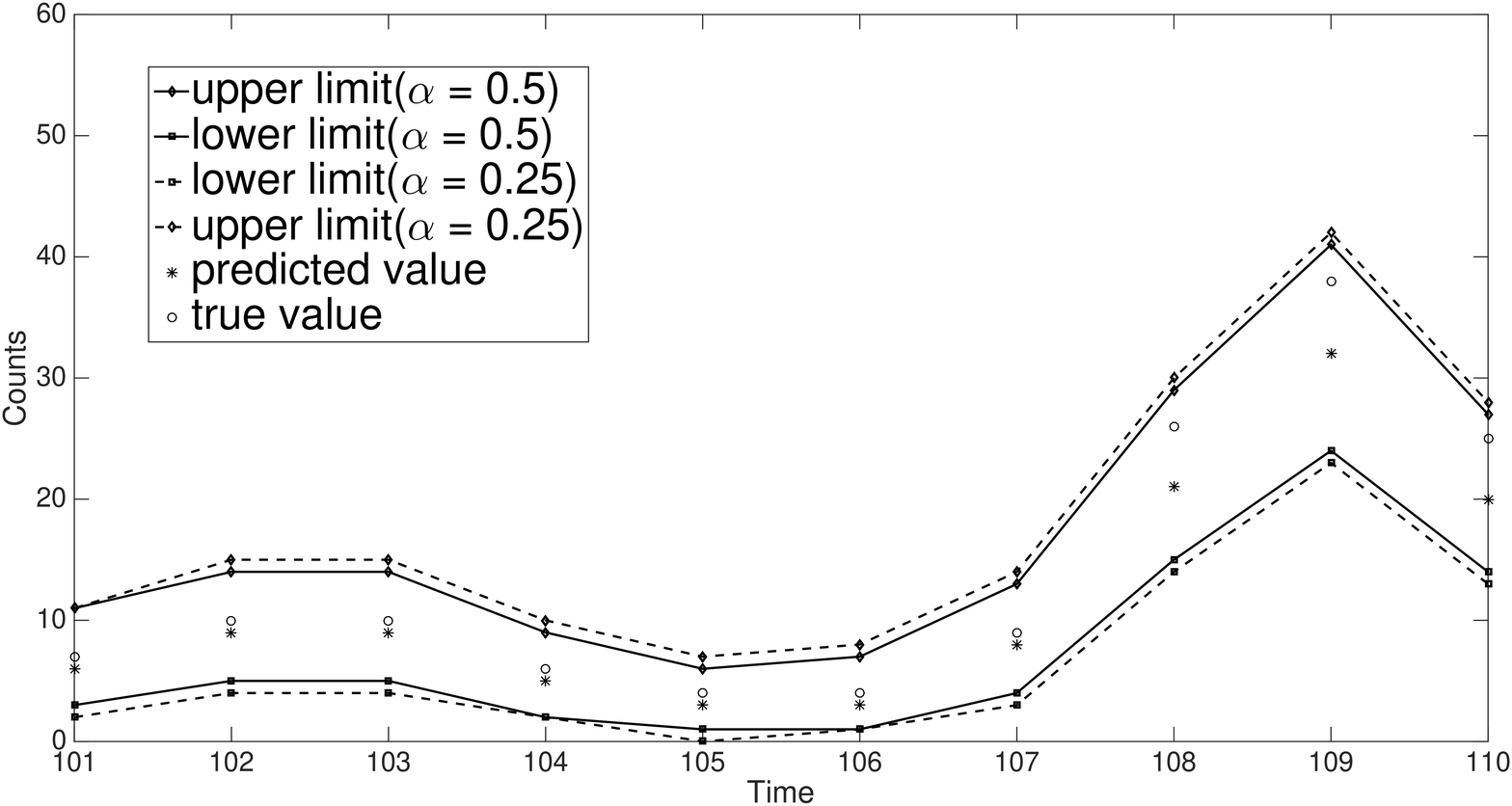}
		\caption{Forecasting results for $y_{101},\ldots,y_{110}$}
		\label{}
	\end{subfigure}
\quad
\vfill
	\begin{subfigure}[b]{0.495\textwidth}
		\centering
		\includegraphics[scale=0.13]{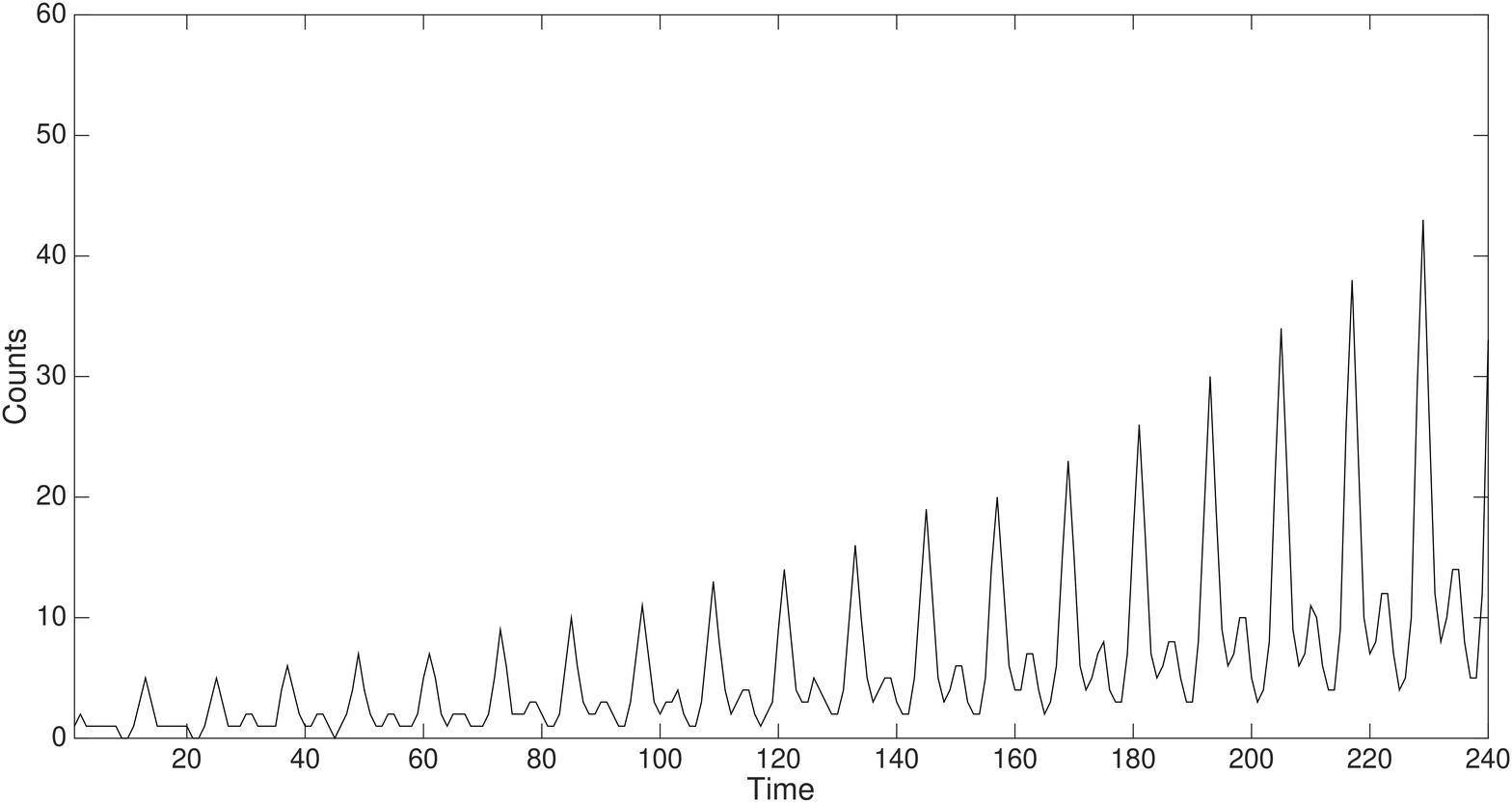}
		\caption{Simulated data, $y_1,\ldots,y_{240}$}
		\label{}
	\end{subfigure}
	\begin{subfigure}[b]{0.495\textwidth}
		\centering
		\includegraphics[scale=0.13]{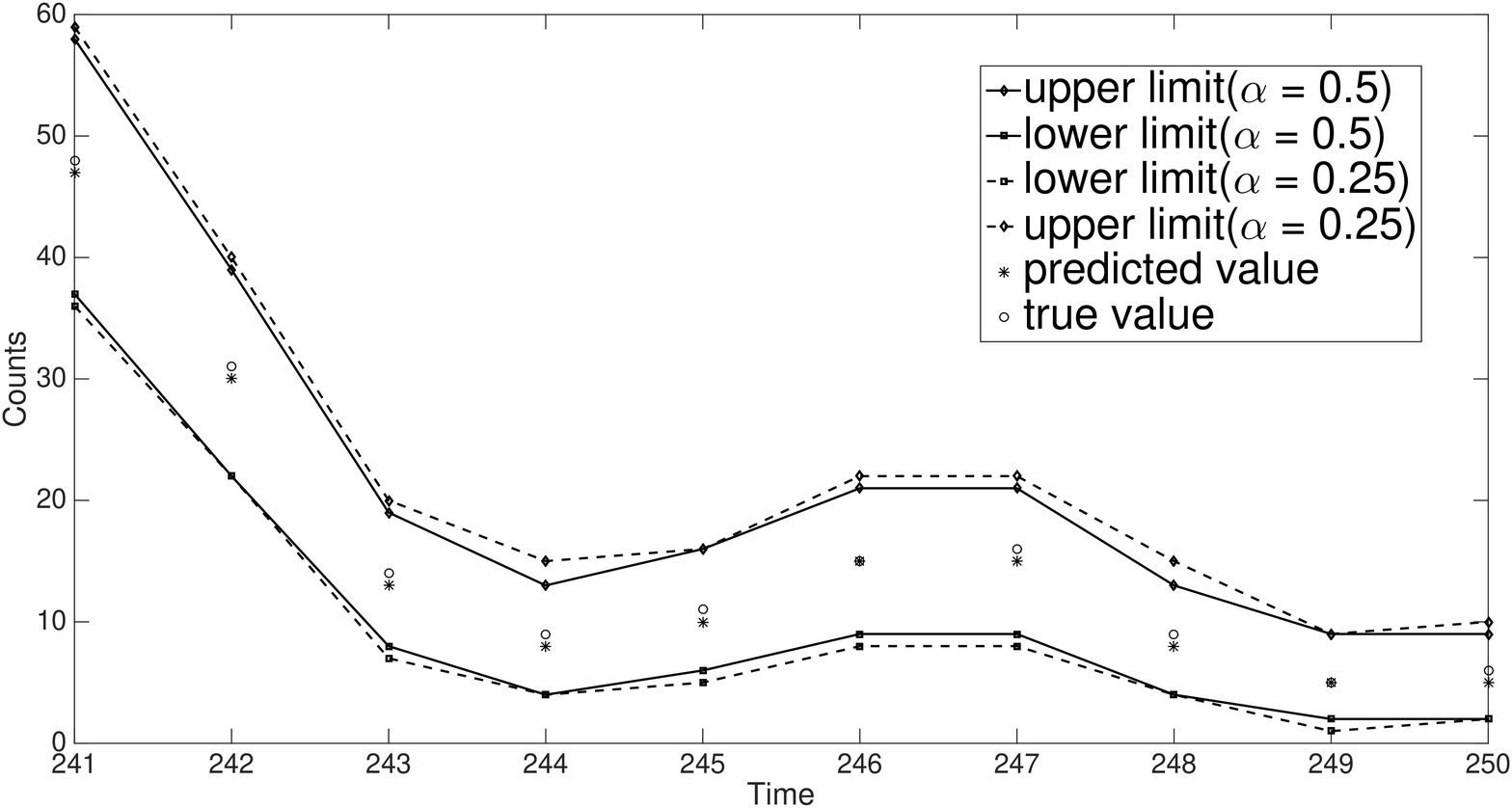}
		\caption{Forecasting results for $y_{241},\ldots,y_{250}$}
		\label{}
	\end{subfigure}
\caption{Simulation and Forecasting results using the PL function for the GARMA $(0, 2)$ model. Note the HDRs of $y_{n+m}$ though shown as continuous intervals contain only integer values.}
\label{}
\end{figure}

 \begin{table}[h!]
	\centering
	\begin{tabular}{|l|l|l|l|l|l|l|}
	
		\hline
		Model       & \multicolumn{3}{c|}{GARMA(0, 2)} & \multicolumn{3}{c|}{GARMA(5, 0)} \\ \hline
		$n$ & 50         & 100      & 240      & 50        & 100        & 240     \\ \hline
		RMSE         & 3.24037      & 3.04959      & 0.89443      & 2.28035     & 2.19089     & 1.89737     \\ \hline
	\end{tabular}
\caption{RMSE values for the two simulation scenarios, GARMA $(0,2)$ and GARMA $(5,0)$  for different values of $n$}
\label{RMSE}
\end{table}

We also assessed the effect of  an empirical probability mass function (pmf) instead of the actual pmf on the HDR computations using a GARMA $(0,2)$ model. The comparisons of the empirical and exact pmfs  for $n=50,100,240$ are shown in Figure 3. We note from Figure 3, as $n$ increases, $\hat{f}_\alpha\rightarrow f_\alpha$, which in turn causes ($\hat{R}(f_\alpha)\rightarrow {R}(f_\alpha)$), i.e., the estimated HDR ($\hat{R}(f_\alpha)$) to approach the true HDR. 

\begin{figure}
	\centering
	\begin{subfigure}[b]{0.495\textwidth}
		\centering
		\includegraphics[scale=0.13]{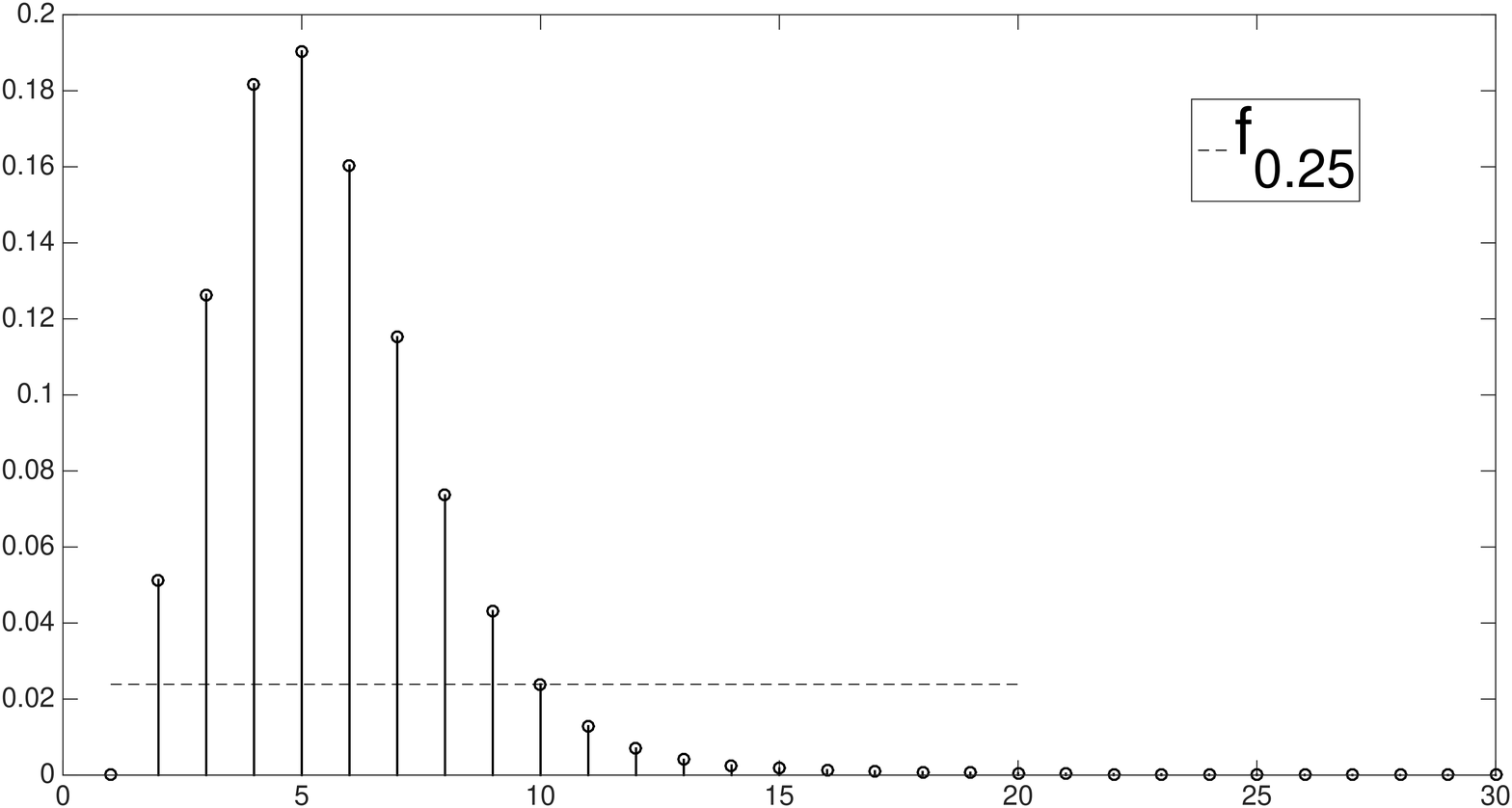}
		\caption{75\% HDR: Exact pmf of $Y_{51}$ }
		\label{}
	\end{subfigure}
	\begin{subfigure}[b]{0.495\textwidth}
		\centering
		\includegraphics[scale=0.13]{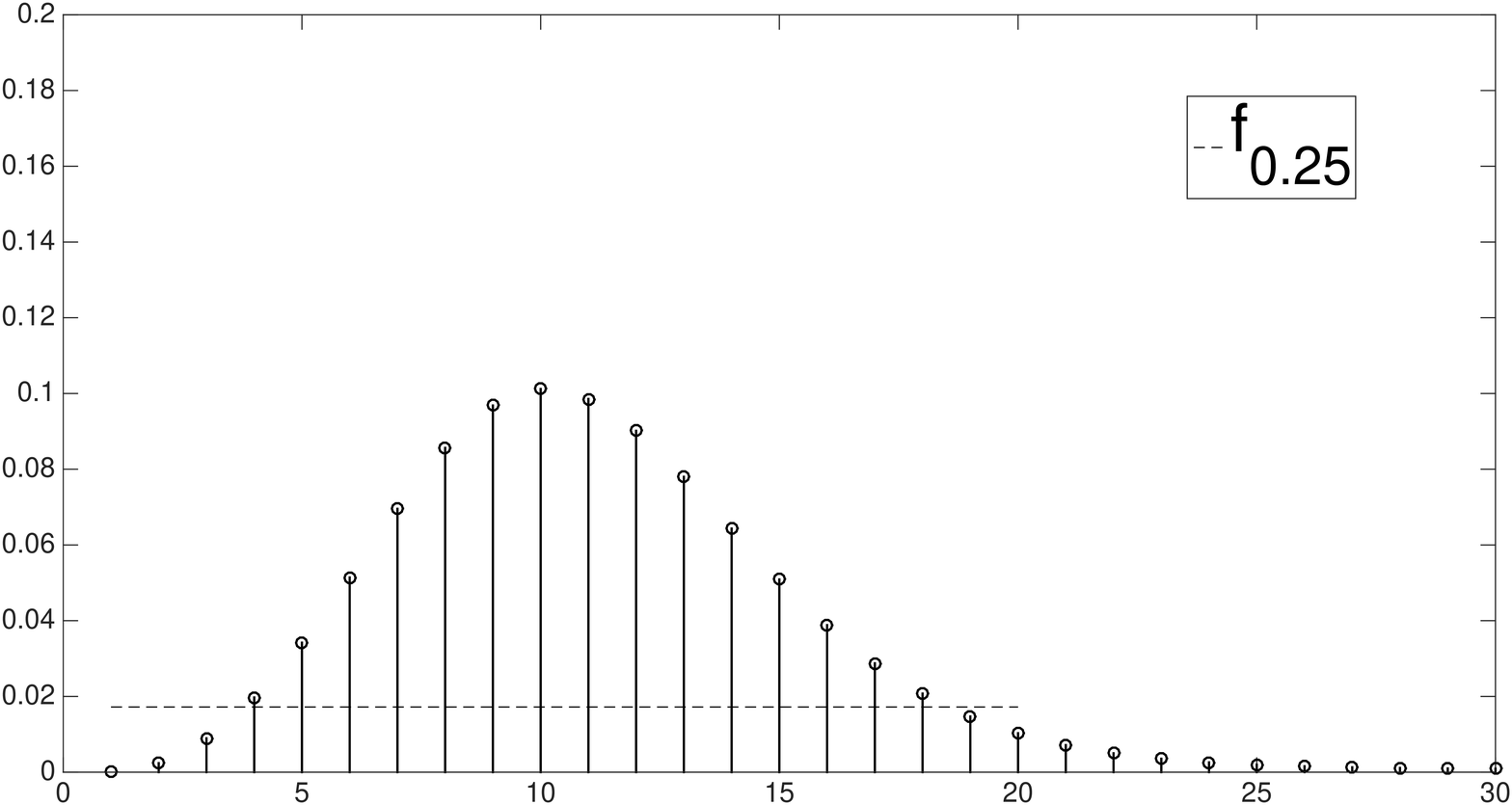}
		\caption{75\% HDR: Empirical pmf of $Y_{51}$}
		\label{}
	\end{subfigure}
\quad
\vfill
	\begin{subfigure}[b]{0.495\textwidth}
	\centering
	\includegraphics[scale=0.13]{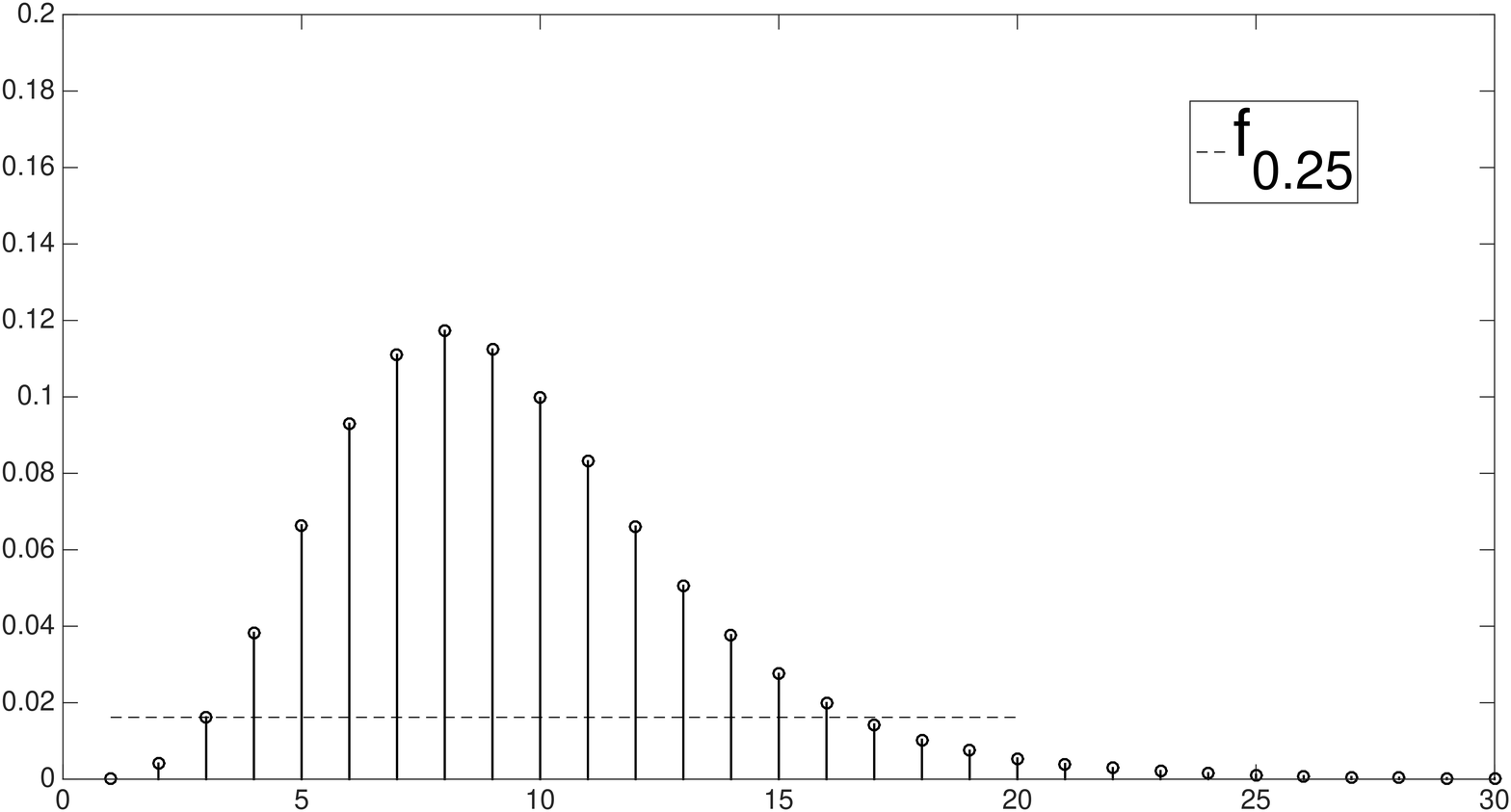}
	\caption{75\% HDR: Exact pmf of $Y_{101}$}
	\label{}
\end{subfigure}
\begin{subfigure}[b]{0.495\textwidth}
	\centering
	\includegraphics[scale=0.13]{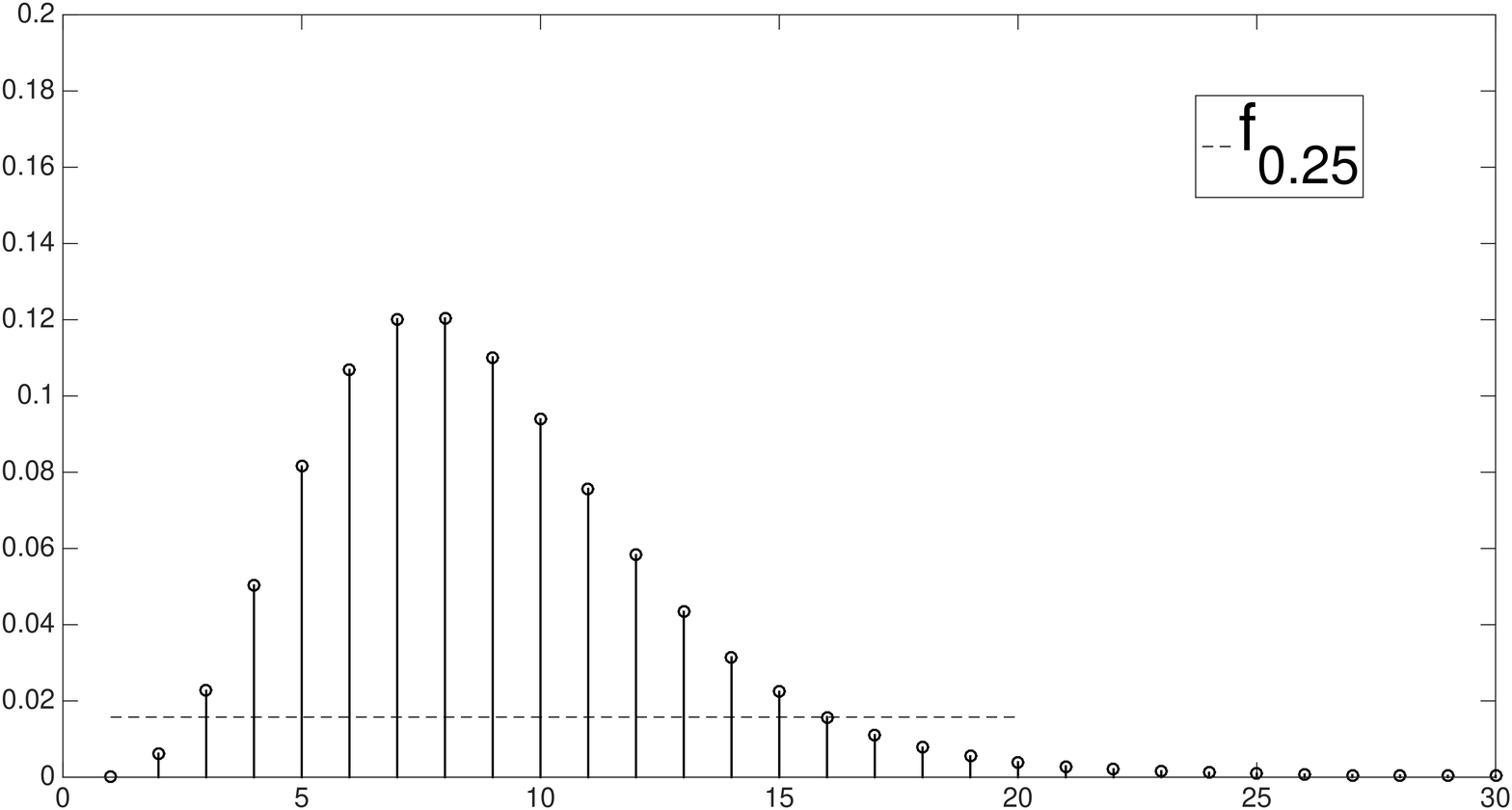}
	\caption{75\% HDR: Empirical pmf of $Y_{101}$}
	\label{}
\end{subfigure}
\quad
\vfill
\begin{subfigure}[b]{0.495\textwidth}
	\centering
	\includegraphics[scale=0.13]{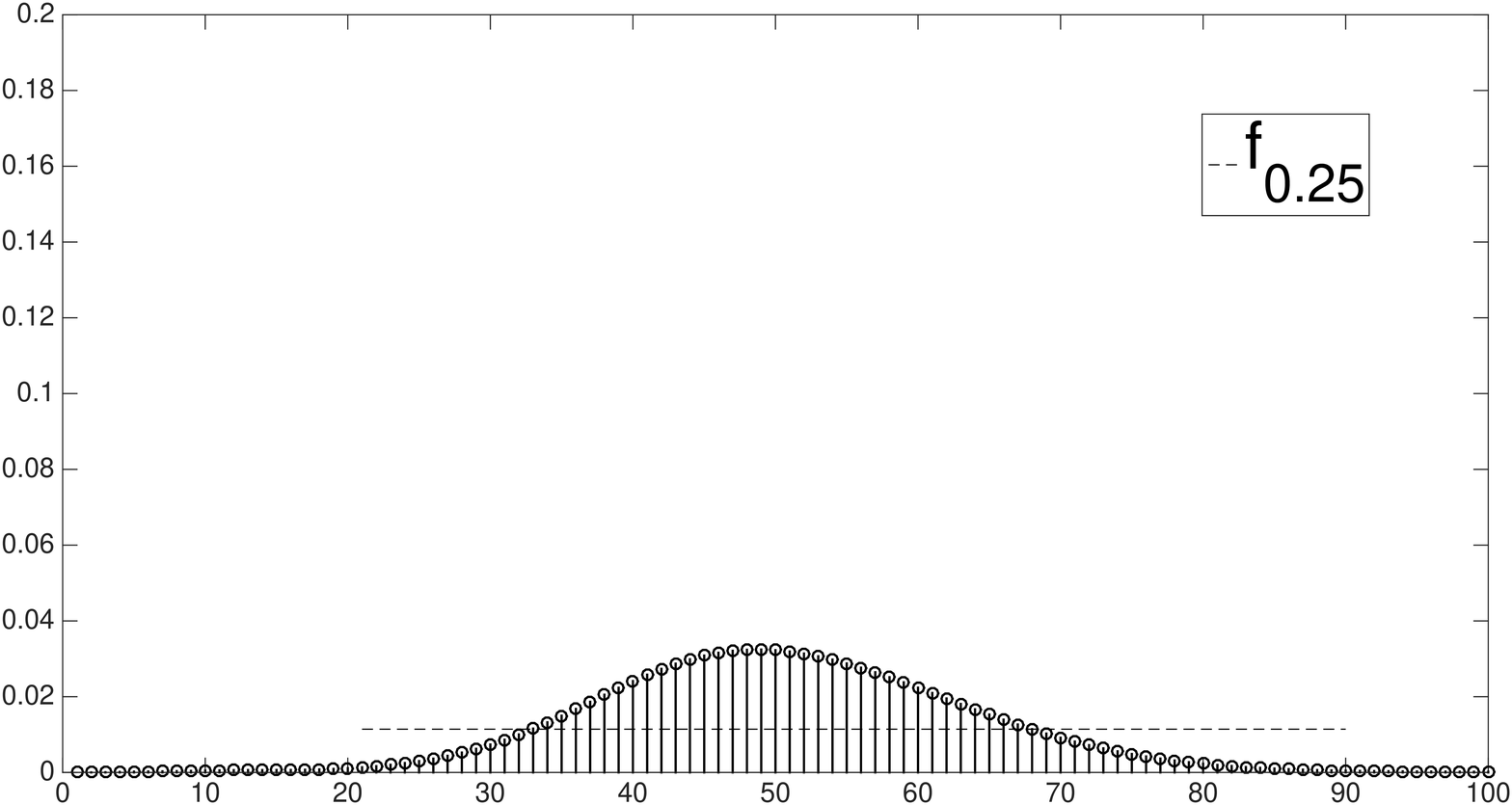}
	\caption{75\% HDR: Exact pmf of $Y_{241}$}
	\label{}
\end{subfigure}
\begin{subfigure}[b]{0.495\textwidth}
	\centering
	\includegraphics[scale=0.13]{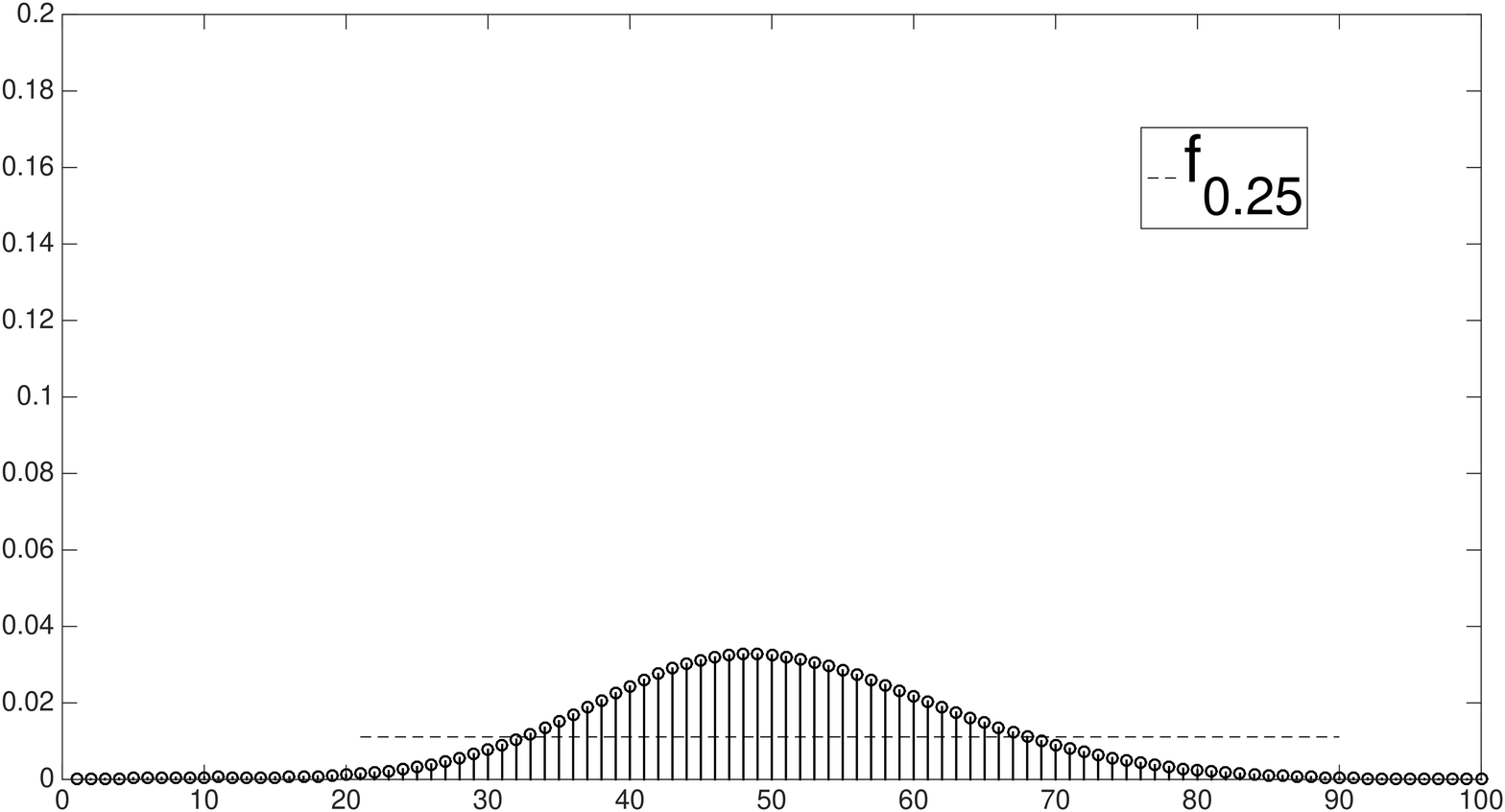}
	\caption{75\% HDR: Empirical pmf of $Y_{241}$}
	\label{}
\end{subfigure}
\caption{Shapes of Exact and Empirical pmfs for different values of $n$ for GARMA $(0,2)$ model}
\label{}
\end{figure}


%
%

\subsection{Forecasting of Polio Data}
To illustrate the new forecasting method proposed we use a  data set  based on monthly cases of poliomyelitis cases reported to the U.S. Centers for Disease
Control for the years 1970 to 1982. This data has been analyzed before by \cite{10.2307/2336303} and \cite{10.2307/30045208}, but no forecasting results have been discussed. 

\begin{table}[h!]
	\centering
	\begin{tabular}{|l|l|l|l|l|l|l|}
		\hline
		& \multicolumn{3}{c|}{\begin{tabular}[c]{@{}l@{}}Parameter Estimates\\   \end{tabular}} & \multicolumn{3}{c|}{\begin{tabular}[c]{@{}l@{}}Standard Errors\\ \end{tabular}} \\ \hline
		Intercept   & \multicolumn{3}{c|}{0.409}                                                             & \multicolumn{3}{c|}{0.122}                                                               \\ \hline
		cos(2pt/12) & \multicolumn{3}{c|}{0.143}                                                             & \multicolumn{3}{c|}{0.157}                                                               \\ \hline
		sin(2pt/12) & \multicolumn{3}{c|}{-0.530}                                                             & \multicolumn{3}{c|}{0.146}                                                               \\ \hline
		cos(2pt/6)  & \multicolumn{3}{c|}{0.462}                                                             & \multicolumn{3}{c|}{0.121}                                                               \\ \hline
		sin(2pt/6)  & \multicolumn{3}{c|}{-0.021}                                                            & \multicolumn{3}{c|}{0.123}                                                               \\ \hline
		MA(1)       & \multicolumn{3}{c|}{0.273}                                                             & \multicolumn{3}{c|}{0.052}                                                               \\ \hline
		MA(2)       & \multicolumn{3}{c|}{0.242}                                                             & \multicolumn{3}{c|}{0.052}                                                               \\ \hline
		Deviance    & \multicolumn{6}{c|}{490.714}                                                                                                                                                     \\ \hline
	\end{tabular}
	\caption{Model estimates for Polio data based on the first 158 observations}
	\label{Estimates and stderr}
\end{table}

We used the first 158 polio cases to fit a Poisson GARMA $(0,2)$ model and then forecasted the  polio counts for years 1972 to 1982. The model fitting statistics along with the estimates and their standard errors are shown in Table \ref{Estimates and stderr}. Figure 4 shows the true counts, predicted counts using the profile likelihood estimation and the $50$ and $75\%$ HDRs. The RMSE value for the 10 forecasts is 1.1186.

\begin{figure}
	\centering
	\begin{subfigure}[b]{0.495\textwidth}
		\centering
		\includegraphics[scale=0.135]{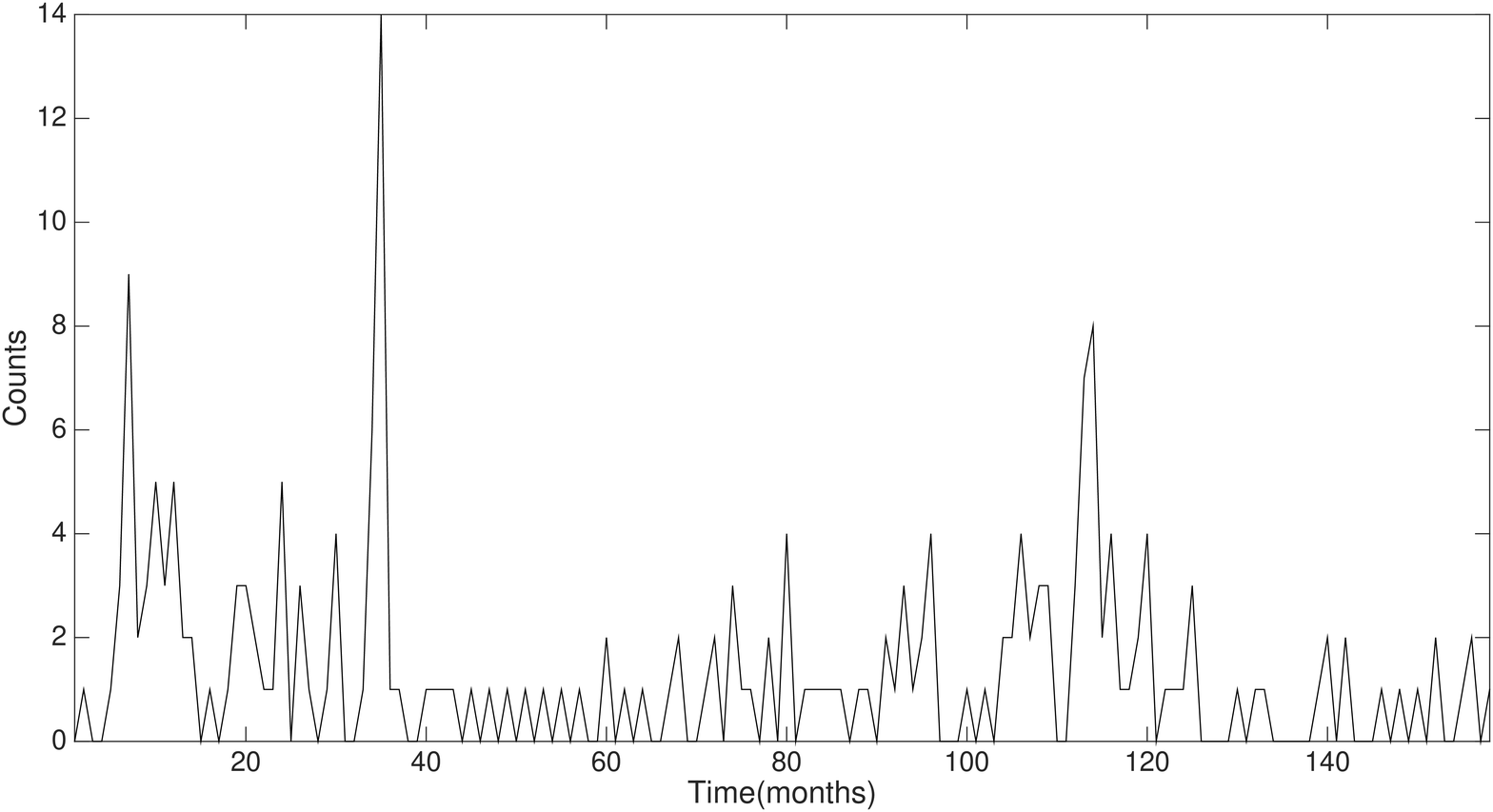}
		\caption{Polio data first 158 obervations}
		\label{}
	\end{subfigure}
	\begin{subfigure}[b]{0.495\textwidth}
		\centering
		\includegraphics[scale=0.135]{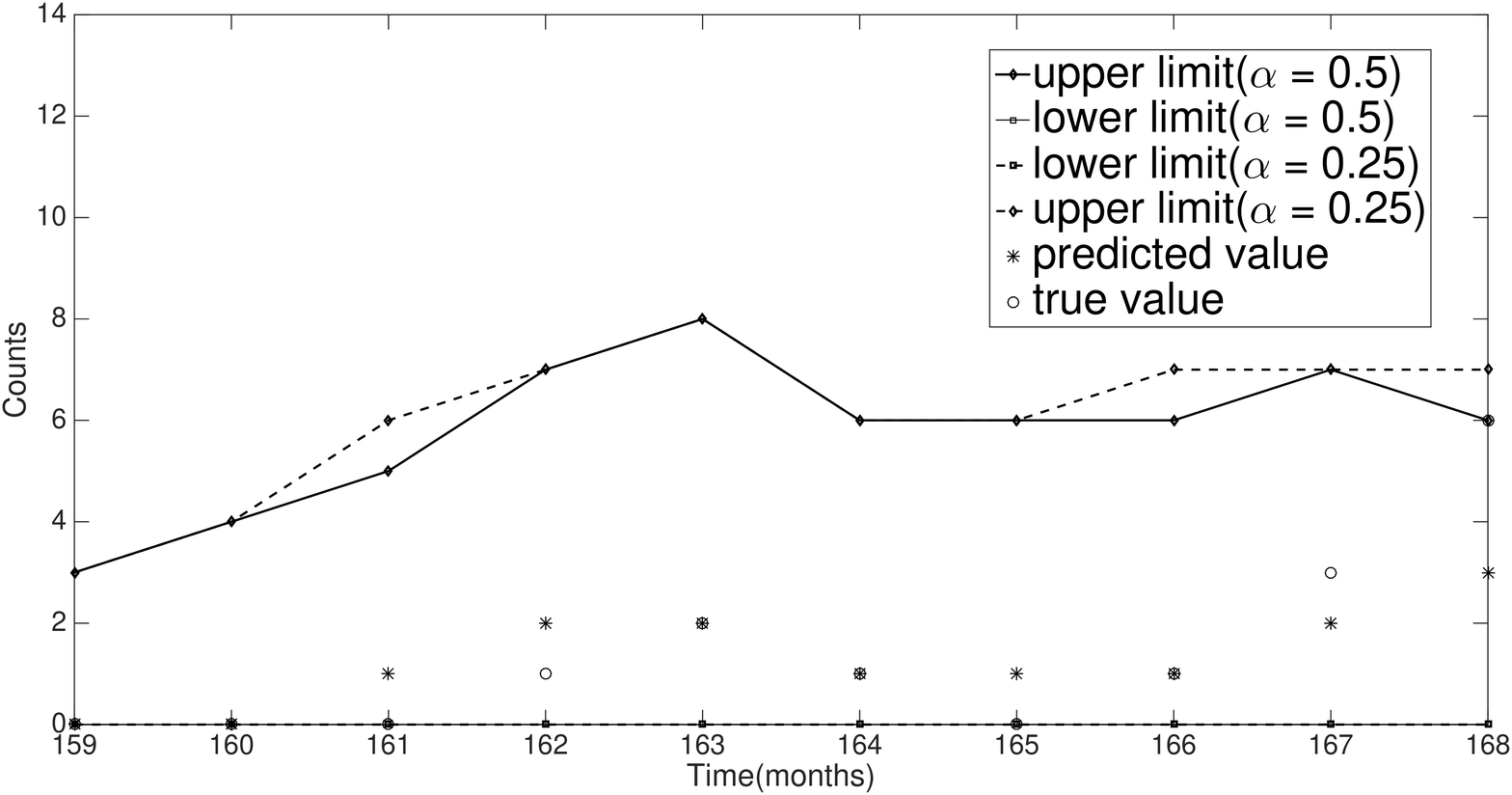}
		\caption{Forecasting results for $y_{159},\ldots,y_{168}$}
		\label{}
	\end{subfigure}
\caption{Forecasting results using the PL function for the Polio data. Note the lower bounds of both HDR regions in Figure 4 (b) are zero. Also, the HDRs of $y_{n+m}$ though shown as continuous intervals, contain only integer values.}
\label{}
\end{figure}

\subsection{Two Step Forecasting using the PL function}
The predictive likelihood function for the two step prediction of the Poisson GARMA$(5,0)$ model  is
  \begin{eqnarray}\label{2step}
 \nonumber {LL}_p(y_{n+2}|H_{n+2})&=&\sum_{y_{n+1}}\frac{\exp(-\sum_{t=1}^{n+2}\hat{\lambda}_t)\prod_{t=1}^{n+2}\hat{\lambda}_t^{y_t}}{\prod_{t=1}^{n+2}y_t!}.  \end{eqnarray} We chose $n=100$.
For predicting $Y_{102}$  we  computed MLEs of the model parameters  based on the observed data and $q$ possible values of the tuplet $\{y_{101},y_{102}\}$. If $\hat{p}(y_{101},y_{102})<1\times 10^{-6}$, then we did not choose that tuplet. The MLEs were then used to find evaluate the normalized ${LL}_p(y_{102}|H_{102})$ for each $y_{102}$ by summing over all  values of $y_{101}$.
The true and forecasted values of ($Y_{102}$)  and the 50 and 75$\%$ HDRs  are reported in Table \ref{twostep}.
\begin{table}[h!]
	\centering
	\begin{tabular}{|l|l|l|l|l|l|}
		\hline
		$Y_{102}$ & $\hat{Y}_{102(PL)}$ & {50$\%$ HDR} &{75$\%$ HDR} \\ \hline
		0         & 0                   &$\{0,1,2\} $           & $\{0,1,2,3,4,5\}$             \\ \hline
	\end{tabular}
\caption{The two step out-of-sample prediction results}
\label{twostep}
\end{table}


\subsection{Robustness Study of the Profile Likelihood Based Prediction}

The PL prediction depends on the model being fitted to the data. In this section, we study the effect of model misspecification on the forecasting. Suppose a data set of size $100$ is drawn from a Poisson GARMA $(0,5)$ model given by,
\begin{eqnarray*}\label{}         
	&\log(\lambda_{t})& =0.2+0.01 t+0.4 \cos(2\pi t/12)+0.5 \sin(2\pi t/12)+0.5\cos(2\pi t/6)\\&&+0.5\sin(2\pi t/6)-0.5\log(y^{*}_{t-1}/\lambda_{t-1})+0.6\log(y^{*}_{t-2}/\lambda_{t-2})+0.01\log(y^{*}_{t-5}/\lambda_{t-5}).
\end{eqnarray*} 
Note the MA coefficients are non zero at lags 1, 2 and 5. However, the coefficient at lag 5 is quite small, so it is easy to misspecify the model as a GARMA (0,2). The estimated GARMA $(0,2)$ model fitted to the data is,
\begin{eqnarray*}\label{}         
	&\log(\lambda_{t})& =0.2285+0.0102 t+0.3488 \cos(2\pi t/12)+0.4995 \sin(2\pi t/12)+0.5225\cos(2\pi t/6)\\&&+0.4371\sin(2\pi t/6)-0.4095\log(y^{*}_{t-1}/\lambda_{t-1})+0.4951\log(y^{*}_{t-2}/\lambda_{t-2}). 
\end{eqnarray*}  
We performed forecasting of $10$ future counts on the basis of the actual and the fitted models and recorded the respective RMSEs, $1.26491$ and $3.03315$.  From Figure 5 we note that the forecasts based on the estimated model are quite close to the true values and also  HDRs successfully captures all the true values except $y_{108}$. Thus, we may conclude on the basis of the example considered that the performance of the PL method in forecasting is not very sensitive to model misspecification. 

\begin{figure}
	\centering
	\begin{subfigure}[b]{0.495\textwidth}
		\centering
		\includegraphics[scale=0.135]{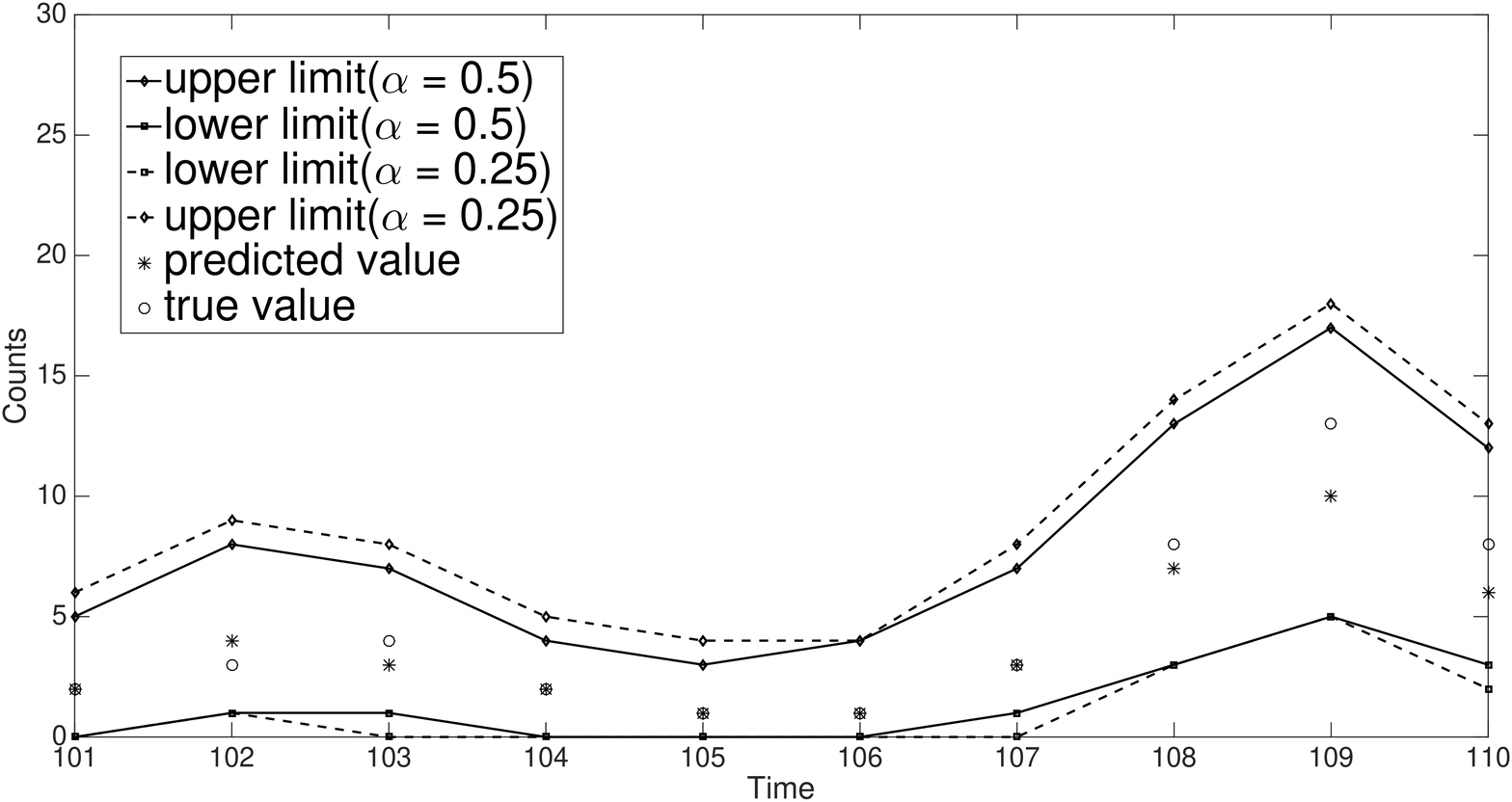}
		\caption{Forecasting results for $y_{101},\ldots,y_{110}$ based on the correct model GARMA $(0, 5)$}
		\label{}
	\end{subfigure}
	\begin{subfigure}[b]{0.495\textwidth}
		\centering
		\includegraphics[scale=0.135]{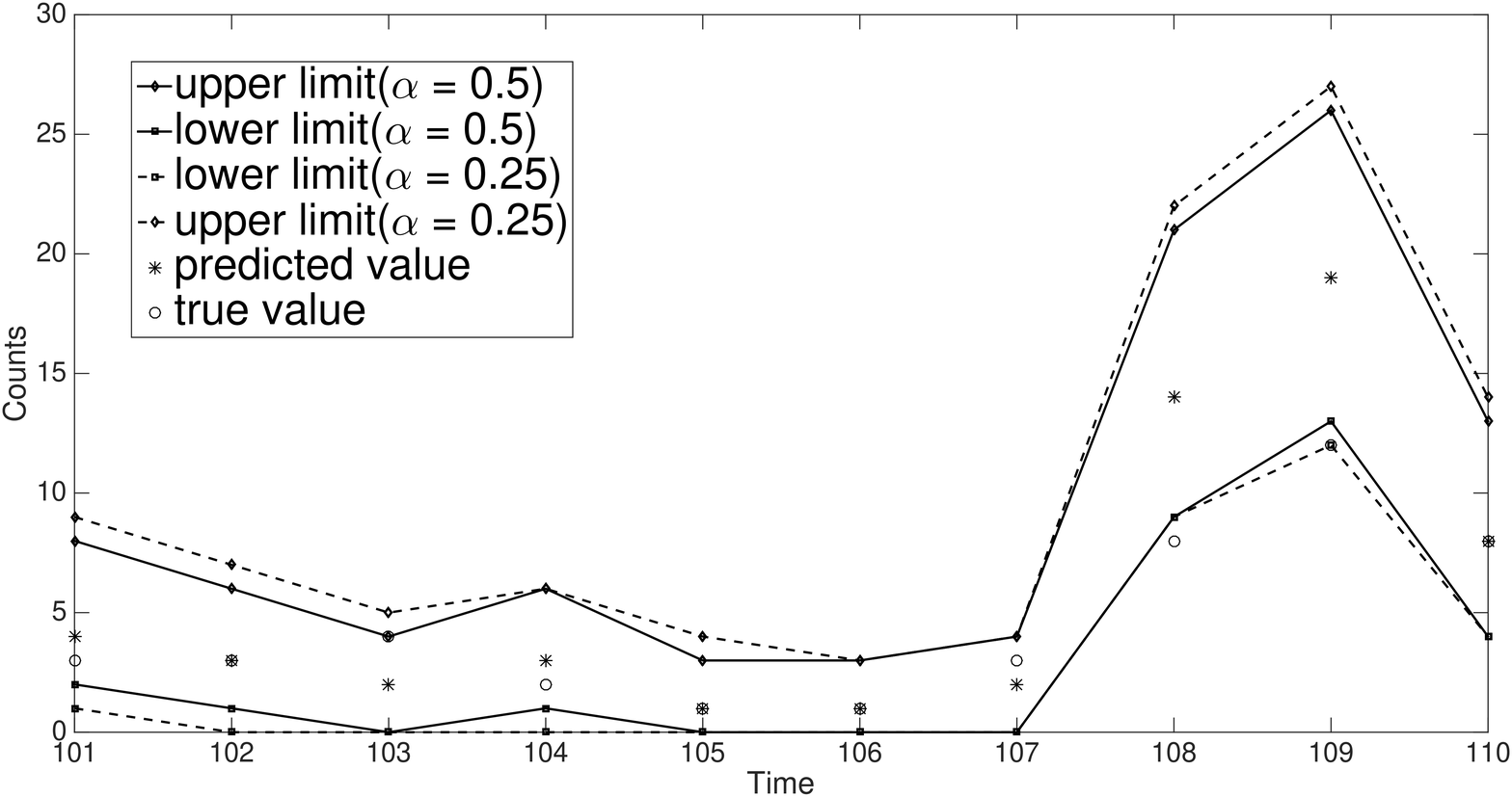}
		\caption{Forecasting results for $y_{101},\ldots,y_{110}$ based on the incorrect model GARMA $(0, 2)$}
		\label{}
	\end{subfigure}
\caption{Robustness results for the PL forecasting method. Note the HDRs of $y_{n+m}$ shown contain only integer values.}
\label{}
\end{figure}

\subsection{Computations}

All computations have been done in Matlab. The GARMA toolbox of \cite{garma} has been used to estimate the model parameters. The time needed to obtain one simulation result for the one step forecast is 49 seconds while for the two step forecast is 404 seconds. The Matlab program for  one step forecast using the PL  function GARMA $(5,0)$ model is given in the supplementary material. Other programs are available from the authors on request.

\section{Concluding Remarks}
In this article we look at several Poisson GARMA time series models for different values of $p$ and $q$, and  use PL functions to find a point forecast and a forecast region. The large sample properties of the estimators based on PL functions are studied. The new forecasting method proposed gives  non negative discrete valued forecasts and forecasting regions coherent with the sample space of the count time series under consideration. 


Very often in practical situations we note that the count data collected usually suffer from overdispersion. In such cases  the negative binomial distribution instead of the Poisson distribution is used. One possible future extension of our work is to use the PL technique to forecast  such overdispersed count time data.

\bibliographystyle{plainnat}
\bibliography{refs}


  \end{document}